%% file: GUT-LHC.tex
\documentclass[a4paper]{JHEP3}
\pdfoutput=1
\usepackage{multirow}
\usepackage{rotating}

\newcommand{\beq}{\begin{equation}}
\newcommand{\eeq}{\end{equation}}
\newcommand{\beqn}{\begin{eqnarray}}
\newcommand{\eeqn}{\end{eqnarray}}

\def\sla#1{\setbox0=\hbox{$#1$}\dimen0=\wd0
      \setbox1=\hbox{/} \dimen1=\wd1 \ifdim\dimen0>\dimen1
      \rlap{\hbox to \dimen0{\hfil/\hfil}} #1                        \else
      \rlap{\hbox to \dimen1{\hfil$#1$\hfil}}
      /   \fi}

\newcommand{\nn}{\nonumber}
\newcommand{\ov}{\overline}

\newcommand{\mtt}{m_{T2}}

\newcommand{\noi}{\noindent}

\newcommand{\Ga}{\Gamma}
\newcommand{\stopi}{ {\tilde t_1} }
\newcommand{\go}{ {\tilde g} }
\newcommand{\chai}{ {\tilde \chi_1^\pm} }
\newcommand{\neui}{ {\tilde \chi_1^0} }
\newcommand{\neuii}{ {\tilde \chi_2^0} }
\newcommand{\grs}{LS{}}
\newcommand{\agrs}{HS{}}

\preprint{TUM-HEP-759/10}

\title{Sparticle masses from transverse mass kinks at the LHC: the case of Yukawa-unified SUSY GUTs}

\author{Kiwoon Choi$^a$, Diego Guadagnoli$^b$, Sang Hui Im$^a$ and Chan Beom Park$^a$\\
$^a$ Department of Physics, Korea Advanced Institute for Science and Technology, Daejon 305-701, Korea\\
$^b$ Excellence Cluster Universe, Technische Universit\"at M\"unchen, Boltzmannstra{\ss}e 2, D-85748 Garching, Germany\\
Email: \email{kchoi@muon.kaist.ac.kr, diego.guadagnoli@ph.tum.de, shim@muon.kaist.ac.kr, cbpark@muon.kaist.ac.kr}\\
(Dated: \today)}

\abstract{ \noi We explore, in a concrete example, to which extent
new particle mass determinations are practicable with LHC data. Our
chosen example is that of Yukawa-unified SUSY GUTs, whose viability
has been recently studied for two general patterns of soft
SUSY-breaking terms.
We note that both patterns of SUSY spectra do not admit long
decay chains, which would make it possible to determine the masses of
the SUSY particles involved using endpoints or mass relations. We
thus take the so-called $m_{T2}$-kink method as our key strategy, since
it does not rely on the presence of long decay chains. 
We then discuss a procedure allowing to determine the masses of the 
gluino, of the lightest chargino as well as of the first two 
neutralinos and, for the scenario where a stop is lighter than the 
gluino, the mass of the light stop too. 
Our worked example of Yukawa-unified SUSY GUTs may offer a useful 
playground for dealing with other theories which predict similar 
patterns of SUSY spectra.}

\begin{document}

\section{Introduction} \label{sec:intro}

\noindent Existing data on established collider quantities --~in
particular electroweak precision tests, quark masses and
flavour-changing neutral current processes~-- provide crucial
constraints on many sets of Standard Model (SM) extensions, whose new 
interactions do in general imply tree- or 
loop-level deviations in some or all of these observables. These
tests allow to learn whether a given class of theories is viable at
all, and, if so, to learn about the features of the viable parameter
space. However these tests, at present, are mostly null tests, 
where the chosen class of theories is required to produce a small 
signal as compared with the SM. It is clear that, in
order to single out the considered class of theories, one has to
find observables for which an unequivocally different behavior with
respect to the SM is predicted. With the imminent flow of direct
data from the Large Hadron Collider (LHC), such observables may be
found in inclusive searches or, better, in the direct determination
of at least the lightest part of the new particles' spectrum. The
second possibility is clearly more ambitious, but also much more
powerful than the first one to tell apart a given theory against
other possibilities. Aim of the present paper is to explore, in a
concrete example, to which extent new particles' mass determinations
are practicable with LHC data.

The example we focus on is that of supersymmetric (SUSY) grand
unified theories (GUTs) with Yukawa unification (YU), whose
viability and expected signatures have been extensively studied with
different approaches for various general patterns of soft
SUSY-breaking terms
\cite{BDR1,BDR2,BF,Tobe:2003bc,Auto:2003ys,BalazsDermisek,Auto:2004km,AABuGuS,Baer:2008jn,AlGuRaS,Baer:2008xc,Baer:2008yd,
Antusch:2008tf,Antusch:2009gu,Gogoladze,GRS,Baer-justso,Baer-Tevatron,Baer-LHC1yr}.%
\footnote{For earlier literature on the topic of YU within SUSY GUTs, the reader is referred to 
\cite{preHRS1,preHRS2,preHRS3,preHRS4,preHRS5,preHRS6,preHRS7,preHRS8,preHRS9,preHRS10,preHRS11,HRS}.}
We will focus here on the two scenarios considered in refs.
\cite{AlGuRaS} and \cite{GRS}, that we briefly summarize in the rest
of this section. In ref. \cite{AlGuRaS} the case where soft-breaking
terms for sfermions and gauginos are universal at the GUT scale was
considered. It was found
\begin{itemize}

\item[{\em (a)}] that this class of theories is phenomenologically viable only by advocating partial decoupling of the
sfermion sector, the lightest mass exceeding 1 TeV;

\item[{\em (b)}] that phenomenological viability can be recovered without decoupling by relaxing $t - b - \tau$ unification
to $b -\tau$ unification, equivalent to a parametric departure of $\tan \beta$ from the value implied by exact YU. This
solution is non-trivial since the constraints from $m_b$ and respectively FCNCs prefer high, O(50), and respectively low
values of $\tan \beta$. Indeed, a compromise solution has been found to exist only for the narrow range
$46 \lesssim \tan \beta \lesssim 48$, implying that the breaking of $t - b$ YU must be {\em moderate}, in the range
10 -- 20\%;

\item[{\em (c)}] that the requirement of $b - \tau$ unification and the FCNC constraints are enough to make
the predictions for the lightest part of the SUSY spectrum robust ones in the interesting region of point {\em (b)}.
These include a lightest stop mass $\gtrsim 800$ GeV, a gluino mass of about 400 GeV and lightest Higgs, neutralino and
chargino close to the lower bounds. Specifically, the lightest neutralino is in the ballpark of 60 GeV, and the lightest
chargino about twice as heavy. The rest of the SUSY spectrum lies instead in the multi-TeV regime: in particular first
and second generation sfermion masses, that almost do not RGE-evolve from the GUT scale to the Fermi scale, remain of the
order of the GUT-scale bilinear mass $m_{16} \gtrsim 7$ TeV. For an illustrative example of this spectrum, see the leftmost
column of table \ref{tab:examplefits}.

\end{itemize}

\begin{table}
\begin{center}
\begin{tabular}{|lc|lc|}
\hline
\multicolumn{4}{|c|}{{\bf Spectrum predictions}} \\
\hline
\multicolumn{2}{|c}{\agrs{} scenario, ref. \cite{AlGuRaS}} & \multicolumn{2}{|c|}{\grs{} scenario, ref. \cite{GRS}} \\
\hline
\hline
$M_{h^0}$  &  121 & $M_{h^0}$  &  126  \\
$M_{H^0}$  &  585 & $M_{H^0}$  &  1109  \\
$M_{A}$  &  586 & $M_{A}$  &  1114  \\
$M_{H^+}$  &  599 & $M_{H^+}$  &  1115  \\
$m_{\tilde t_1}$  &  783 & $M_{\tilde t_1}$  &  192 \\
$m_{\tilde t_2}$  &  1728 & $m_{\tilde t_2}$  &  2656 \\
$m_{\tilde b_1}$  &  1695 & $m_{\tilde b_1}$  &  2634 \\
$m_{\tilde b_2}$  &  2378 & $m_{\tilde b_2}$  &  3759 \\
$m_{\tilde \tau_1}$  &  3297 & $m_{\tilde \tau_1}$  &  3489  \\
$m_{\tilde\chi^0_1}$  &  59 & $m_{\tilde\chi^0_1}$  &  53  \\
$m_{\tilde\chi^0_2}$  &  118 & $m_{\tilde\chi^0_2}$  &  104  \\
$m_{\tilde\chi^+_1}$  &  117 & $m_{\tilde\chi^+_1}$  &  104  \\
$M_{\tilde g}$  &  470 & $M_{\tilde g}$  &  399  \\
\hline
\end{tabular}
\caption{Spectrum predictions for representative fits of the viable scenarios studied in refs. \cite{AlGuRaS} and \cite{GRS}.
All masses are in units of GeV. Uppercase and lowercase masses stand for pole and respectively $\overline{\rm DR}$ masses.}
\label{tab:examplefits}
\end{center}
\end{table}
Ref. \cite{GRS}, instead, explored the complementary case where YU is kept exact, while the requirement of universal soft terms
at the GUT scale (not very justified from a theoretical standpoint) is relaxed. Ref. \cite{GRS} focused on the scenario in which the breaking
of universality inherits from the Yukawa couplings, i.e. is of minimally flavor violating type \cite{MFV,Snowmass96}. This general setup allows
in particular for a splitting between the up-type and the down-type soft trilinear couplings. Comparison with data revealed for
this trilinear splitting scenario the following features \cite{GRS}:
\begin{itemize}

\item[{\em (i)}] agreement with data clearly selects the parametric region with a large $\mu$-term and a sizable splitting between
the $A$-terms. At the price of a slight increase in the fine tuning required to achieve EWSB with precisely the correct value of
$M_Z$, this scenario allows a substantial improvement on other observables that, on a model-dependent basis, do often require some
amount of fine tuning as well;

\item[{\em (ii)}] in particular, and quite remarkably, phenomenological viability does not invoke a partial decoupling of
the sparticle spectrum, as is the case in the scenario of ref. \cite{AlGuRaS}, but instead it {\em requires} various SUSY particles,
notably the lightest stop and the gluino, to be very close to their current experimental limits. Specifically, the lightest stop is
lighter than the gluino, in the range [100, 200] GeV. This testable difference reflects the substantial difference in the mechanism
that makes the two classes of models phenomenologically viable. Ultimately, it maps onto the difference in the assumed form for the
GUT-scale soft terms;

\item[{\em (iii)}] concerning the rest of the spectrum, the predictions for the gluino, the lightest chargino and neutralino masses,
as well as those for the heavy part of the SUSY spectrum, are very similar to the corresponding ones in the scenario of \cite{AlGuRaS}
(see ref. \cite{GRS} for further details). Again, because of the cross-fire between the many constraints used, all these spectrum
predictions, including those of point {\em (ii)}, are robust ones.

\end{itemize}

As already mentioned, this paper is devoted to exploring the possibility of directly measuring the masses of the lightest SUSY particles
in the scenarios of refs. \cite{AlGuRaS,GRS}. In particular, it will discuss a strategy --~applicable already to data
collected at the LHC~-- that permits the determination of the masses of the gluino, the lightest chargino and neutralino and,
for the scenario of ref. \cite{GRS}, the lightest stop also.
This task clearly requires to focus on {\em short} decay chains, and makes extensive use of so-called ``transverse mass'' variables.

Our strategy will be described in section \ref{sec:strategy}, after an introduction to the methods for new particles' mass
determination at hadron colliders, to be presented in the next section.

\section{Methods for SUSY mass determinations at the LHC} \label{sec:methods}

In this section we briefly recall the methods devised so far for
new-particles' mass determinations at the LHC, and discuss their relation with
our problem of interest. A comprehensive and
very recent review can be found in ref. \cite{BarrLester-review}.

\subsection[Methods other than $\mtt$]{\boldmath Methods other than $\mtt$}

The most straightforward mass-determination method is, in general,
to find a peak in the invariant mass distribution of the decay
products of the new particle of interest. This is however
inapplicable if the final state includes a component that escapes
detection, which is our case, because of the missing `lightest SUSY particle' (LSP).

For decay chains that include LSPs, in order to reconstruct SUSY
masses one can exploit kinematic relations involving these masses
and the (measured) momenta of the visible particles. Among the
strategies devised to this end, a first possibility is given by the
so-called {\em endpoint method} \cite{Hinchliffe1,Bachacou1,
Hinchliffe2,Allanach1,Gjelsten1,Weiglein1,Gjelsten2,
Lester+,Gjelsten3}.
It is based on the
observation that, in a given decay chain, the endpoint values of the
invariant mass distributions constructed for visible decay products
depend on the masses of the invisible particles as well. A prototype
example is provided by the squark decay chain \cite{ATLAS-DTP}
\beqn
\label{eq:endpoint-example}
\tilde q ~\to~ \tilde \chi^0_2 \, q
~\to~ \tilde \ell^\pm \ell^\mp q ~\to~ \tilde \chi^0_1 \ell^+ \ell^-
q~,
\eeqn
which would be available for
$m_{\tilde{q}}>m_{\tilde\chi_2^0}>m_{\tilde{\ell}}>m_{\tilde\chi_1^0}$.
Assuming that the four-momenta of the leptons and of the $q$-initiated jet
can all be determined with reasonable accuracy, the above decay
process provides four invariant-mass distributions, i.e. those of
$m_{\ell \ell}$, $m_{q \ell ({\rm high})}={\rm
max}(m_{q\ell^+},m_{q\ell^-})$, $m_{q \ell ({\rm low})}={\rm
min}(m_{q\ell^+},m_{q\ell^-})$, and $m_{q \ell \ell}$. Then their
endpoints can be inverted to give all the four sparticle masses
involved in the decay.
However this decay mode is kinematically closed for our scenarios of
interest, since all the sleptons are (much) heavier than any of the
other sparticles involved in eq. (\ref{eq:endpoint-example}).

Besides, it can be shown that, in order for all the sparticle masses
to be, even in principle, {\em separately} reconstructible, the
endpoint  method requires necessarily a long decay chain with at
least 3 decay steps \cite{Burns-massrel}. If this condition is not
fulfilled, the number of measurable invariant masses is smaller than
the number of unknown particle masses, and one can at best determine
some combinations of mass differences, rather than determining the
absolute mass values. A simple example would be a single-step 3-body
decay
\beqn \label{chi02decay}
\tilde \chi^0_2 ~\to~ \tilde \chi^0_1 \ell^+ \ell^-~.
\eeqn
In this case, the dilepton invariant mass has an endpoint equal to
$m_{\tilde\chi_2^0}-m_{\tilde\chi_1^0}$.
In order to determine $m_{\tilde\chi_2^0}$ and $m_{\tilde\chi_1^0}$
separately, one would need additional kinematic variables as discussed
in \cite{ATLAS-DTP}.

A second method is that of {\em mass relations}\footnote{This method
is often called the polynomial method.}
\cite{Nojiri1,massrel1,Cheng1,massrel2,massrel3}. The idea here is
that the four-momenta of missing particles in the decay products can 
be reconstructed by exploiting various constraints
including the on-shell constraints on the decay chain. Provided the
number of branches in the decay chain is sufficiently high, the
number of constraints can be made to exceed the number of unknowns,
and one can solve (or actually fit) for all the unknowns (for a
detailed discussion see \cite{Burns-massrel}). For example, in an
$n$-step cascade decay initiated by a SUSY particle $Y$, with each
vertex branching into a SUSY particle $I_i$ and a SM visible
particle with (reconstructible) momentum $p_i$, 
\beqn 
Y \rightarrow I_1+V(p_1) \rightarrow I_2+V(p_2)+V(p_1) 
\rightarrow ... \rightarrow I_n(k)+V(p_n)+...+V(p_1)~, 
\eeqn 
one has $n+1$ independent mass-shell constraints, namely 
\beqn 
k^2 = m_\chi^2~, (k+p_n)^2 = m^2_{I_{n-1}}~, ...~, 
(k+p_1+...+p_n)^2 = m_Y^2~, 
\eeqn 
where $k$ is the four-momentum of the missing particle $I_n$. 
For $N$ such events, the number of constraints will be $N(n+1)$, 
while the number of unknowns will instead be $4 N + (n+1)$, i.e. 
the four-momentum of the missing particle in each event and the $n+1$
masses of the SUSY particles which are common to all events. In
order for the number of constraints to exceed the number of
unknowns, one needs $n\geq 4$ and also $N\geq (n+1)/(n-3)$.

A case where the mass-relation method would be more appropriate is
that of symmetric $n$-step decays of pair-produced SUSY particles,
$Y+\bar{Y}$, for which one can use the missing transverse momentum
constraint also \cite{Cheng1,massrel3}. In this case, again for $N$
events, the unknowns include as before $n+1$ SUSY particle masses
which are common to all events, and the four-momenta of the two
missing particles in each event, so the total number of unknowns is
given by $(n+1)+8N$. As for the available constraints, for each
event one has $2(n+1)$ constraints from the mass-shell conditions
and two constraints from the missing transverse momentum, so
$2N(n+1)+2N$ constraints in total. We then find the number of
constraints is equal to or bigger than the number of unknowns if
$n\geq 3$ and $N\geq (n+1)/2(n-2)$. These observations lead to the
conclusion that one needs at least a 3-step or a 4-step cascade
decay in order for the mass-relation method to be applicable.

On the other hand, for the scenarios of refs. \cite{AlGuRaS,GRS} 
the predicted SUSY spectra imply that, at the 
energies available at the LHC, there cannot be any long decay chain 
($n\geq 3$) on which the above discussed endpoint or mass-relation 
methods might realistically be applied.
Specifically, the only 3-step cascade decay with sizable number of 
events is the gluino decay in the scenario of \cite{GRS}, 
$\tilde{g} \rightarrow t\tilde{t}_1^* \rightarrow b W \bar{b} \tilde{\chi}_1^-
\rightarrow b q q \, \bar{b} q q \tilde{\chi}_0$, 
which however suffers from very large jet combinatorics. As a result, the 
mass-relation method simply cannot be used in our case, and the endpoint 
method can determine at best mass differences. 

\subsection[The $\mtt$-kink method]{\boldmath The $\mtt$-kink method}

A third method exists, however, that is able to
determine SUSY masses even in the absence of long decay chains. This
method, called the {\em $\mtt$-kink method}
\cite{kink1,Gripaios1,Barr-kink,kink2,Nojiri-kink,kink3,cheng-han},\footnote{%
For another method, also applicable to short decay chains, see 
\cite{mt2-decomposition}.}
exploits the fact that the endpoint value of the transverse mass
variable $\mtt$, regarded as a function of the trial mass
$m_\chi$ of the missing particle in the decay products, exhibits a
kink at $m_\chi=m_\chi^{\rm true}$ \cite{kink1}. As the endpoint
value of $\mtt$ at $m_\chi=m_\chi^{\rm true}$ corresponds to the
mother particle mass, the $\mtt$-kink method determines both
the mother particle mass and the missing particle mass
simultaneously. As we will see, applying the $\mtt$-kink method to
pair-produced gluinos in the scenario of \cite{GRS}, while
regarding charginos as missing particles, one can determine both the
gluino mass and the chargino mass. Likewise, for the scenario of
\cite{AlGuRaS}, one can consider the $\mtt$-kink of gluino pairs
while regarding the second lightest neutralinos as missing
particles.
With those masses determined by the $m_{T2}$-kink, the rest of the 
light part of the SUSY spectrum can be determined by applying the 
endpoint method.

The collider observable $\mtt$ is a generalization of the transverse
mass $m_T$ which has been introduced for decay processes producing
invisible particles \cite{LesterSummers,BarrLesterStephens}.
Specifically, considering the decay $Y\rightarrow V(p)+\chi(k)$,
where $\chi(k)$ is an invisible particle with four-momentum
$k$, and $V(p)$ stands for an arbitrary number of visible particles
with total four-momentum $p$, the transverse mass $m_T$ is defined
as 
\beqn \label{MTdef} 
m_T^2 ~=~ m_V^2 + m_\chi^2 + 2 \left( \sqrt{m_V^2 + |\vec p_T|^2} 
\sqrt{m_\chi^2 + |\vec k_T|^2} - \vec p_T \cdot \vec k_T \right),~ 
\eeqn 
where $m_V$ and $m_\chi$ are the
invariant masses of $V$ and $\chi$, respectively. As the
four-momentum of $\chi$ cannot be directly measured, let $k$ and 
$m_\chi^2 \equiv k^2$ denote a trial four-momentum and respectively 
a trial mass squared for $\chi$, which are introduced for the
sake of discussion. A key feature of $m_T$ is that it is bounded by
the invariant mass, i.e. the true mother particle mass $m^{\rm
true}_Y$, when $m_\chi$ and $\vec k_T$ take the true values: 
\beqn
&& m_T^2(m_\chi=m_\chi^{\rm true},\,\vec{k}_T =
\vec{k}_T^{\rm true}) \,\leq\, (m_Y^{\rm true})^2 \nonumber \\
&=&m_V^2+(m_\chi^{\rm true})^2 + 2 \left( \sqrt{m_V^2 + |\vec p_T|^2}
\sqrt{(m_\chi^{\rm true})^2 + |\vec k^{\rm true}_T|^2}\cosh
(\eta_V-\eta_\chi^{\rm true}) - \vec p_T \cdot \vec k^{\rm true}_T
\right)\nonumber 
\eeqn 
\FIGURE[ht]{
  \includegraphics[width=0.4\textwidth]{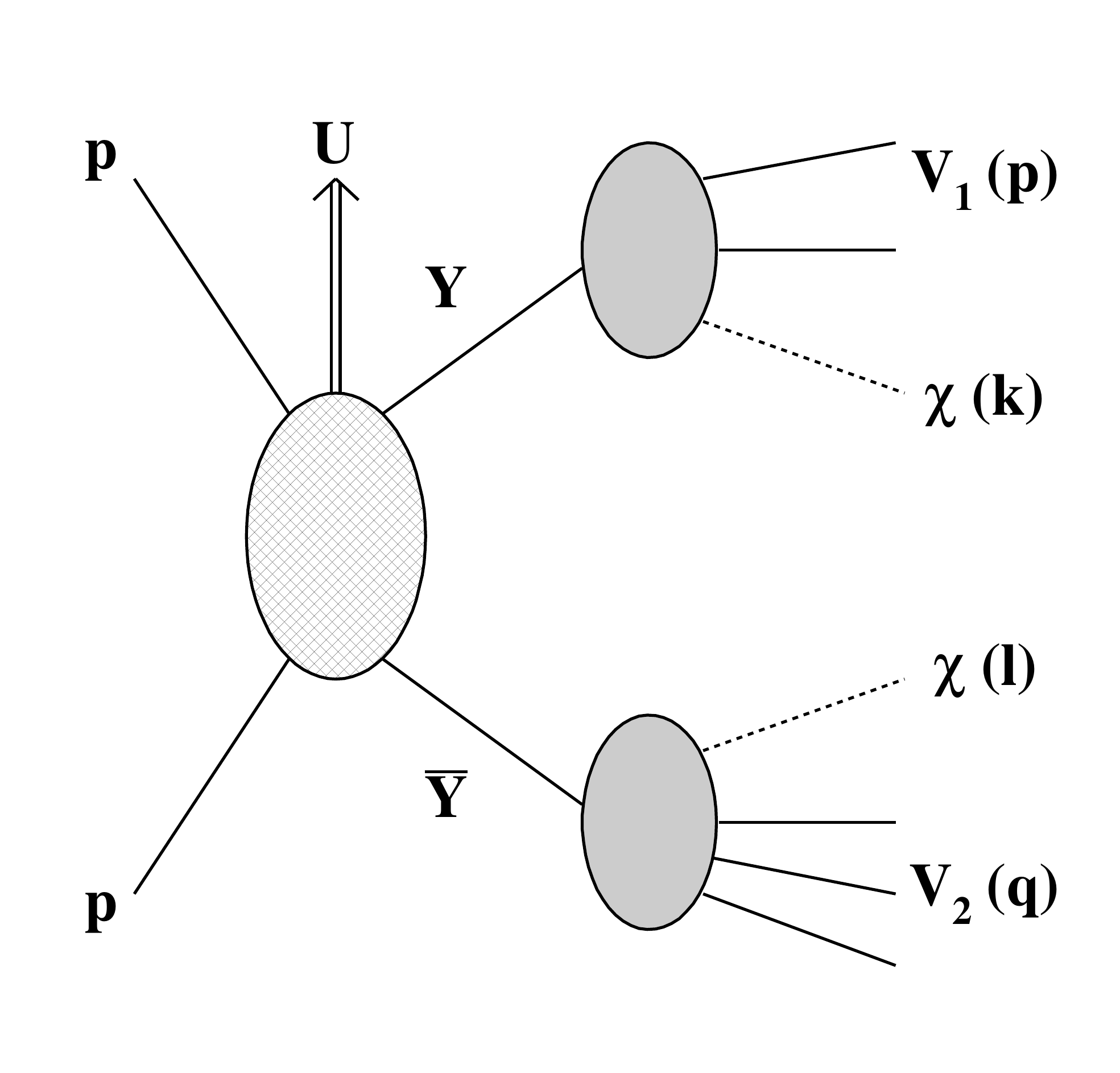}
  \caption{Sketch of event topology relevant for the applicability of the
  $\mtt$-kink method.}
  \label{fig:MT2event}
}
where $\eta=\frac{1}{2}\ln \left(\frac{E+p_z}{E-p_z}\right)$ is the 
pseudo-rapidity. While the true momentum of $\chi$ cannot be directly 
measured, its transverse
component $\vec k^{\rm true}_T$ can be inferred from the missing
transverse momentum if $\chi$ is the only missing particle in the
whole event: $\vec k^{\rm true}_T=\sla{\vec p}_T$. Then, once
$m_\chi^{\rm true}$ is known by some other information, one can
determine $m^{\rm true}_Y$ from the endpoint value of
$m_T(m_\chi=m_\chi^{\rm true},\vec{k}_T=\vec{k}_T^{\rm
true}=\sla{\vec p}_T)$.

As typical SUSY events involve a pair of SUSY particles decaying
into two invisible LSPs, the transverse mass should be accordingly
generalized. At the LHC, generic SUSY events take the form (see fig.
\ref{fig:MT2event})
\beqn
p + p &\to& U+Y + \ov Y \\
&\to& U+V_1(p) + \chi(k) + V_2(q) + \chi(l)~, \nn 
\eeqn 
where $U$, 
often called the upstream momentum, denotes the total four-momentum
of visible particles not coming from the decays of the SUSY particle
pair $Y+\ov Y$, $\chi$ is the invisible LSP, and again each $V_i$
($i=1,2$) denotes an arbitrary number of visible particles produced
by the decay of $Y$ or of $\ov Y$. (The upstream momentum might come
from initial or final state radiation or from the decay of a heavier
SUSY particle.) Unlike the case of single missing particle, the
transverse momentum of each LSP cannot be determined in this case,
although their sum can be read off from the missing transverse
momentum. One can then introduce trial LSP transverse momenta,
$\vec{k}_T$ and $\vec{l}_T$, under the constraint
$\vec{k}_T+\vec{l}_T=\sla{\vec p}_T$, and consider the transverse
mass of each decay mode. Then the $\mtt$ variable for a trial LSP
mass $m_\chi$ can be constructed as 
\cite{LesterSummers,BarrLesterStephens} 
\beqn 
\mtt({\rm event}; m_\chi) \equiv {\rm min}_{\vec k_T + \vec l_T = 
\sla{\vec p}_T} \left[ {\rm max} \left( m_T(Y\rightarrow V_1+\chi),
~m_T(\ov Y\rightarrow V_2+\chi) \right) \right]~, 
\eeqn 
where each event is
specified  by the set of {\em measured} kinematic variables
$\{m_{V_1}, \vec p_T, m_{V_2}, \vec q_T, \sla{\vec p}_T \}$, with
$\sla{\vec p}_T$ the total missing transverse momentum. Note that,
due to the relation $\sla{\vec p}_T = -\vec{p}_T -\vec{q}_T -
\vec{U}_T$, use of $\sla{\vec p}_T$ as an event variable is
equivalent to using the upstream transverse momentum $\vec{U}_T$.
One can now observe two main properties of the $\mtt$ variable:
({\em a}) once the rest of the input is fixed (i.e. given the
event), $\mtt$ is a monotonic (increasing) function of the trial LSP
mass; ({\em b}) when the trial LSP mass is equal to the true LSP
mass, $\mtt$ is bounded from above by the true mother-particle mass
$m_Y^{\rm true}$.

It follows that, studying $\mtt$ as a function of $m_\chi$ in the
neighborhood of $m_\chi=m_\chi^{\rm true}$, and recalling that $\mtt
(m_\chi=m_\chi^{\rm true})\leq m_Y^{\rm true}$, one will observe a
kink structure at the point $\{\mtt, m_\chi\}=\{m_Y^{\rm true},
m_{\chi}^{\rm true}\}$, because the $\mtt$ curves generally feature
different slopes at $m_\chi=m_\chi^{\rm true}$. See fig.
\ref{fig:MT2kink} for an illustration. In fact, a relevant issue for
the $\mtt$-kink method is how sharp the kink is, which is equivalent
to how much the slopes of $\mtt$ curves (at $m_\chi=m_\chi^{\rm
true}$) vary over the available event set. It has been noted that
endpoint events with different $m_V$ or different $\vec{U}_T$
generically have different slopes
\cite{kink1,Gripaios1,Barr-kink,kink2,Nojiri-kink,kink3}.
\FIGURE[h!]{
 \includegraphics[width=0.4\textwidth]{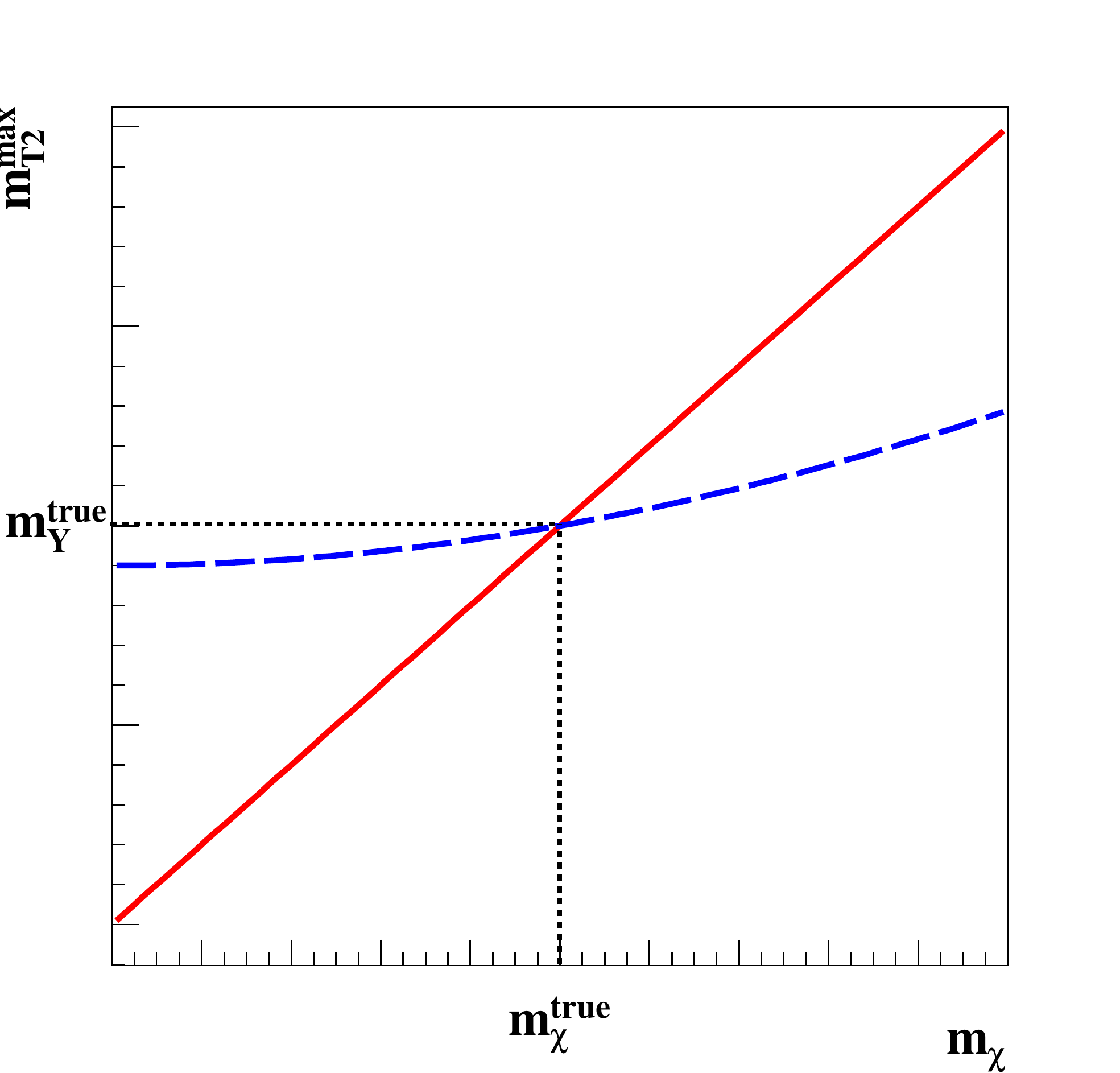}
 \caption{Pictorial representation of the $m_{T2}$ curve
for two events (solid vs. broken lines) in the endpoint region close to
$\mtt (m_\chi=m_\chi^{\rm true})=m_Y^{\rm true}$. The $x$- and $y$-axes
are in arbitrary units.}
 \label{fig:MT2kink}
} 
\noindent This implies that the kink can be sharp enough to be
visible if $V$ corresponds to multi-particle states with wide range
of $m_V$ and/or if a large $|\vec U_T|$ is available. As the
possible range of $m_V$ is bounded by $m_Y^{\rm true}-m_\chi^{\rm
true}$, the kink can be sharp enough only when $m_Y^{\rm true}$ is
substantially  heavier than $m_\chi^{\rm true}$. As for $|\vec
U_T|$, although in principle it can be substantially larger than
$m_Y^{\rm true}$, in reality the best possible situation would be
$|\vec U_T| \sim m_Y^{\rm true}$.

\subsection{Generalizations}

Although $\chi$ was identified as the LSP in the above
discussion, the very same argument applies to the more general case 
where $\chi$ is a generic SUSY particle produced by the decay of $Y$ or
$\bar{Y}$, as long as $\chi(k)$ and $\chi(l)$ have the same mass \cite{Nojiri_subsys,Burns-massrel}.
For instance, $\chi$ might be a heavier neutralino or a chargino,
decaying into the LSP and some SM particles, provided one regards all
those particles as missing particles. Indeed, as already mentioned, in 
the following we will be applying the $\mtt$-kink method to gluino-pair 
decay in the scenarios of \cite{AlGuRaS,GRS}, with $\chi$ being a chargino 
or the second lightest neutralino respectively.\footnote{%
One can further generalize $\mtt$ to the case in which the parent or daughter
particles are not identical \cite{BGL-mt2generalization,Konar-mt2generalization}.} 
Looking at the SUSY spectra, 
summarized in table \ref{tab:examplefits}, the gluino undergoes a relatively
short (2-step) cascade decay into $\chi$ plus several visible particles, 
and also the gluino mass is sensibly above
$m_\chi$. This by itself indicates that a visible kink might be
possible even with events having a negligible $|\vec U_T|$.

In cases where the visible decay products involve multiple jets, 
it is often difficult to identify the endpoint position because of
the smearing caused by poor jet momentum resolution. In such cases, to
reduce the endpoint smearing, one can consider an alternative
form of $\mtt$ defined {\em solely} in terms of the jet {\em transverse}
momenta \cite{kink1,kink2}: 
\beqn 
\mtt ={\rm min}_{\vec k_T + \vec l_T = \sla{\vec p}_T} \left[ {\rm max} 
\left( m_T(Y\rightarrow V_1+\chi),
~m_T(\ov Y\rightarrow V_2+\chi) \right) \right]~, 
\eeqn
where now $m_T(Y\rightarrow V+\chi)$ is given by
\beqn \label{MTdef1} 
m_T^2 ~=~ m_{TV}^2 + m_\chi^2 + 2 \left( \sqrt{m_{TV}^2 + |\vec p_T|^2} 
\sqrt{m_\chi^2 + |\vec k_T|^2} - \vec p_T \cdot \vec k_T \right).~ 
\eeqn 
Here
\beqn \label{MTdef2}
m_{TV}^2(Y\rightarrow V+\chi) =\sum_\alpha m_\alpha^2+2\sum_{\alpha>\beta}
(E_{\alpha T}E_{\beta T}-\vec{p}_{\alpha T}\cdot \vec{p}_{\beta T})~,
\eeqn 
where $\vec{p}_{\alpha T}$ is the measured transverse momentum of the
$\alpha$-th jet in the set of visible decay products $V$, and
$E_{\alpha T}=\sqrt{m_\alpha^2+|\vec{p}_{\alpha T}|^2}$, with $m_\alpha$ 
denoting the $\alpha$-th {\it parton} mass, rather than the jet invariant 
mass, i.e. $m_\alpha=0$ for non-$b$ jets and $m_\alpha=m_b$ for $b$-tagged 
jets.
As the transverse momentum can be measured with better accuracy
than the other components, the $\mtt$ defined in this way suffers
less from jet momentum uncertainty.
In addition, it still obeys all the features for a kink at 
$m_\chi = m_\chi^{\rm true}$ with $\mtt^{\rm max}(m_\chi=m_\chi^{\rm true})
= m_Y$ at partonic level. On the other
hand, as $m_{TV}\leq m_V$, the $\mtt$ defined with the transverse mass 
$m_{TV}$ has generically smaller statistics near the endpoint than the 
$\mtt$ defined with the full invariant mass $m_V$. Previous
studies suggest that the reduced statistics is not a severe drawback
compared to the gain from reduced  smearing \cite{kink1,kink2}, and
thus the $\mtt$ defined in terms of the jet transverse momenta can
reveal a kink even when the kink of $\mtt$ defined with $m_V$ is
smeared away due to poor jet momentum resolution. In the following
analysis including detector effects, we will use the $\mtt$
defined with $m_V$ if its distribution shows a clear endpoint, and 
the $\mtt$ defined with $m_{TV}$ otherwise.

To conclude this section, we would like to mention that several new 
methods of mass measurement for events involving invisible particles 
have recently been proposed, e.g. one exploiting
the cusp structure of the distribution of certain kinematic
variables \cite{iwkim1} and another based on a generic algebraic
singularity that arises in the observable phase space obtained by
projecting out the unmeasurable kinematic variables \cite{iwkim2}.
It would be worthwhile to investigate how much useful those 
methods may be for the specific SUSY scenarios of \cite{AlGuRaS,GRS} 
that we are studying in this paper.

\section{Strategy} \label{sec:strategy}

In this section we apply the idea of $\mtt$-kink  described above to
the SUSY GUT scenarios exemplified in table \ref{tab:examplefits},
and henceforth referred to as \agrs{} (heavy stop) scenario (ref.
\cite{AlGuRaS}) and \grs{} (light stop) scenario (ref. \cite{GRS}),
because of the heavy and respectively light stop relatively to each
other.
\input Tables/table_SUSYprod.tex
\input Tables/table_decays.tex
A first necessary piece of information is that of the main SUSY production
cross sections for the two scenarios. Table \ref{tab:cross-sections}
shows the most important production mechanisms, for $p p$ events with $\sqrt s$
equal to 10 or 14 TeV.%
\footnote{Our analysis will be focused on the 14 TeV case. The 10 TeV case is
reported for guidance, with the aim of showing that production cross sections
for all considered processes are such that our strategy should still apply at
this value of $\sqrt s$.}
They have been calculated with Pythia 6.42 \cite{Pythia6.4}.
For both \agrs{} and \grs{} scenarios the dominant SUSY-production mechanism
is $\tilde g \tilde g$. Interestingly, for the \grs{} scenario, $\tilde t_1 \tilde t_1^*$
production is also large, close to 40\%. Finally, also $\tilde \chi_2^0 \tilde \chi_1^\pm$
associated production is non-negligible in both scenarios.
The second needed piece of information is that of the main decay modes for
the produced particles. These decay modes are reported in table \ref{tab:decay-modes},
and calculated with SUSYHIT \cite{SUSYHIT}. We will elaborate on these production and decay
figures in due course during the analysis.

In the light of the information in tables \ref{tab:cross-sections} and \ref{tab:decay-modes},
the rest of this section is devoted to a short description of our mass-determination
procedure, as it would be carried out in a parton-level analysis. In order to provide
a quick overview of the various steps, the full procedure is also schematically reported
in table \ref{tab:strategyrecap}. This table shows that our procedure is able
to determine the $\tilde g$, $\tilde \chi^0_1$, $\tilde \chi^0_2$, $\tilde \chi_1^\pm$
masses in either scenario and, for the \grs{} scenario, also the $\tilde t_1$ mass.
The concrete implementation of our strategy, carried out on 100/fb of LHC data at the
design center-of-mass energy of 14 TeV, and including realistic detector effects, is
postponed to section \ref{sec:results}.

\newcommand{\GRSgogo}{1}
\newcommand{\GRSgogoTEXT}%
{{\normalfont \itshape Construct $\mtt$ for $\tilde g \tilde g \to$ 4 $b$ + 2 $W$ (+ 2 $\ell$) + $\sla{p}_T$
decay events, thereby measuring $m_{\tilde g}$ and $m_\chai$}}
\newcommand{\GRStt}{2}
\newcommand{\GRSttTEXT}%
{{\normalfont \itshape For $\tilde t_1 \tilde t_1 \to$ 2 $b$ + 4 $q$ + $\sla{p}_T$ events,
determine the mass differences $m_\stopi - m_\neui$ and $m_\chai - m_\neui$ from the
endpoints of the transverse-mass distributions $m_{T,bqq}$ and $m_{T,qq}$, respectively}}
\newcommand{\GRSchaneulll}{3}
\newcommand{\GRSchaneulllTEXT}%
{{\normalfont \itshape For $\chai \neuii \to$ $\ell^+ \ell^-$ + $\ell'$ + $\sla p_T$ events,
determine the mass difference $m_\neuii - m_\neui$ from the endpoint of the invariant-mass
distribution $m_{\ell \ell}$}}
\newcommand{\AGRSgogo}{1}
\newcommand{\AGRSgogoTEXT}%
{{\normalfont \itshape Construct the $\mtt$ variable for $\tilde g \tilde g \to$ 4 $b$
(+ 2 $\gamma$) + $\sla p_T$ decay events, thereby measuring $m_{\tilde g}$ and $m_\neuii$}}
\newcommand{\AGRSchaneulll}{2}
\newcommand{\AGRSchaneulllTEXT}{\GRSchaneulllTEXT}
\newcommand{\AGRSchaneuqqll}{3}
\newcommand{\AGRSchaneuqqllTEXT}%
{{\normalfont \itshape For $\chai \neuii \to$ 2$q$ + 2$\ell$ + $\sla{p}_T$ events, determine
the mass difference $m_\chai - m_\neui$ from the endpoint of the invariant-mass distribution
$m_{qq}$ (and again $m_\neuii - m_\neui$ from $m_{\ell \ell}$)}}

\input Tables/table_strategyrecap.tex

\subsection{Strategy: \grs{} scenario}

Given the generally lighter masses and the higher cross sections for the \grs{} scenario,
our strategy is richer in this case, and we describe it first.

\FIGURE[ht]{
\includegraphics[width= 0.45 \textwidth]{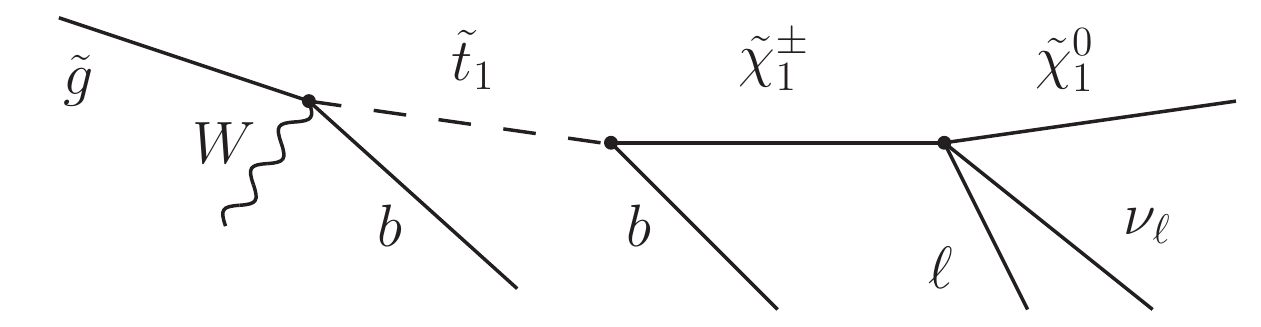}
\caption{Decay chain of interest for step \GRSgogo{} (\grs{} scenario).}
\label{fig:GRSgogo}
}
\subsubsection*{Step \GRSgogo{}: \GRSgogoTEXT}

\noindent In the scenario of ref. \cite{GRS}, the gluino decays
always as $\go \to \stopi t$, with the top quark subsequently
decaying as $t \to b W$ and the $\stopi$ as $\chai b$. Requiring the
$\chai$ to decay as $q q' \neui$ would have the advantage that all
the decay products apart from the $\neui$ would be reconstructible,
{\em in principle}. However, in practice, the event trigger would
consist of 2 $b$ jets and 4 additional non-$b$ jets (2 jets from $W$
and another 2 jets from $\tilde{\chi}_1^\pm$) for {\em each} of the
two decay chains, making this channel very challenging already on
account of the jet combinatoric error.

A more promising strategy is to focus instead on
leptonically-decaying charginos, namely $\chai \to \neui \ell
\nu_\ell$, and to regard the whole $\chai$-initiated decay chain
(see fig. \ref{fig:GRSgogo}) as $\sla{p}_T$. In this case the final
state will contain $2 b + W + \ell \nu_\ell + \neui$ for each decay
chain, with $\ell$ either of $e, \mu$. The $\sla{p}_T$ will
then be the sum of the transverse momenta of the $\neui$ and of the
$\ell \nu$ pair. Concerning the $W$, one can require it to decay in
a $q q'$ pair, so that $m_{q q} = m_W$ (up to the $W$ width) in each
of the two decay chains. The event topology to look at is therefore
$4 b + 2 W + 2 \ell + \sla{p}_T$: while the combinatoric problem has
not completely disappeared, it has been substantially mildened  with
respect to the case mentioned in the previous paragraph.

Constructing the $\mtt$-kink  for
this event topology (with, as mentioned, the 2 leptons' $p_T$
included in $\sla p_T$) allows to determine simultaneously $m_\go$
and $m_\chai$.


\FIGURE[ht]{
\includegraphics[width=0.36 \textwidth]{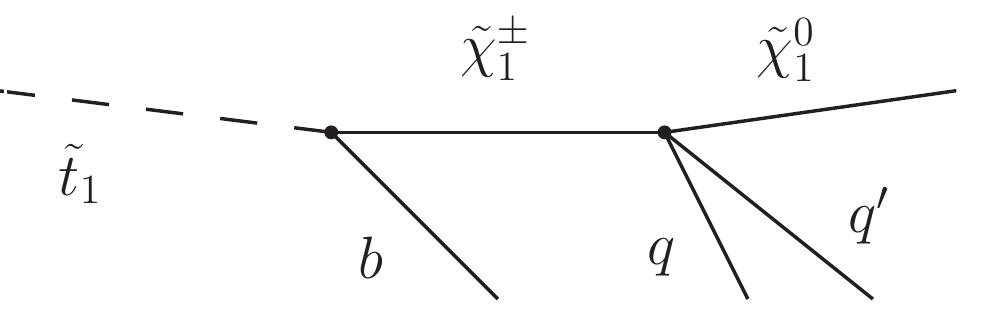}
\caption{Decay chain of interest for step \GRStt{} (\grs{}
scenario).} \label{fig:GRStt} }
\subsubsection*{Step \GRStt{}: \GRSttTEXT}

\noindent In the scenario of ref. \cite{GRS}, given the lightness of the $\stopi$,
its pair-production is substantial and the only kinematically allowed decay mode is
$\stopi \to \chai b$. One can further require the $\chai$ to decay (besides the invisible
$\neui$) into visible products (see fig. \ref{fig:GRStt}), which happens in about
20\% of the cases. The mass differences $m_\stopi - m_\neui$ and $m_\chai - m_\neui$
can then be obtained as the endpoints of the transverse-mass distributions calculated
respectively on the $b q q'$ and $q q'$ systems, for each of the two decay chains of
the event. An evident problem here is to correctly assign the $b$- and $q$-jets to
the two decay chains. We will come back to this in the analysis.

\subsubsection*{Step \GRSchaneulll{}: \GRSchaneulllTEXT}

\FIGURE[hb]{
\includegraphics[width= 0.38 \textwidth]{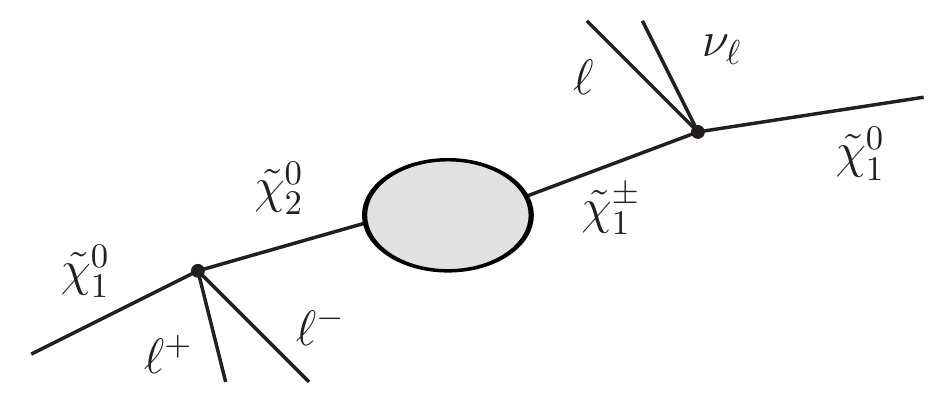}
\caption{Decay chain of interest for step \GRSchaneulll{} (\grs{} scenario).}
\label{fig:GRSchaneu}
}
\noindent Within the \grs{} scenario, information on the $m_\neuii$ mass can only be
obtained from direct production channels, as a simple inspection of table
\ref{tab:decay-modes} reveals. In particular, one can consider the quite large
$\chai \neuii$ associated production, and require both sparticles to decay to
leptons (besides the $\neui$), namely $\neuii \to \neui \, \ell^+ \ell^-$ and
$\chai \to \neui \, \ell \nu_\ell$ (see fig. \ref{fig:GRSchaneu}).
Reconstructing the visible part of the $\neuii$ decay allows to determine the mass
difference $m_\neuii - m_\neui$, hence $m_\neuii$, since the lightest neutralino mass
is already known from the previous step \GRStt{}.

\subsection{Strategy: \agrs{} scenario}

\noindent We next turn to discussing SUSY mass determinations within the \agrs{} scenario
of table \ref{tab:examplefits}.

\FIGURE[h]{
\includegraphics[width= 0.35 \textwidth]{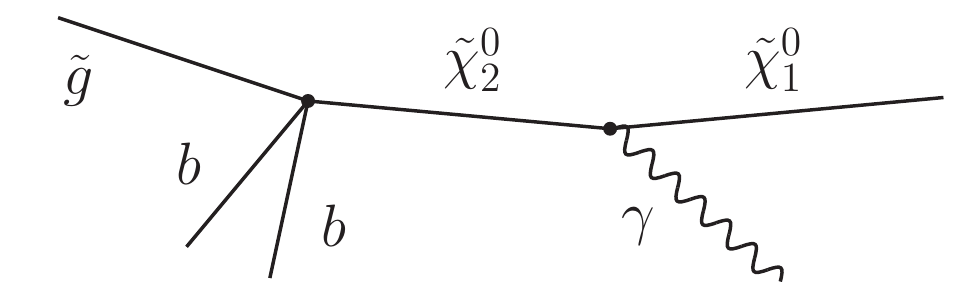}
\caption{Decay chain of interest for step \AGRSgogo{} (\agrs{} scenario).} 
\label{fig:AGRSgogo} } 
\subsubsection*{Step \AGRSgogo{}: \AGRSgogoTEXT}

\noindent Gluino-gluino
production, followed by gluino decaying as $\go \to b \ov b \neuii$,
is probably the golden mode for the \agrs{} scenario. If one further
considers the radiative decay of the $\neuii$ into $\neui$ (note that 
this decay mode has the largest branching ratio), the
implied event trigger, namely $4 b + 2 \gamma + \sla p_T$, has very
small contamination from SM background. Including the gammas in the
$\sla p_T$, the kink in the $\mtt$ variable constructed for these
events allows to simultaneously determine $m_\go$ and $m_\neuii$.

Alternatively, one might attempt to apply the $\mtt$-kink method
to the subsystem involving $2\neuii$ as parent particles and $2\gamma$
as their visible decay products. In this case, the kink in $\mtt$
arises due to nonzero upstream transverse momentum and it allows to 
determine the masses of $\neuii$ and $\neui$. From the analysis we
find however that the resulting kink is not as clean as the one obtained
in the approach of the previous paragraph, where also the $4b$ are included
as visible decay products.

\subsubsection*{Step \AGRSchaneulll{}: \AGRSchaneulllTEXT}

The implementation of this step is entirely analogous as in step \GRSchaneulll{} of the \grs{}
scenario.

\FIGURE[h]{
\includegraphics[width=0.39 \textwidth]{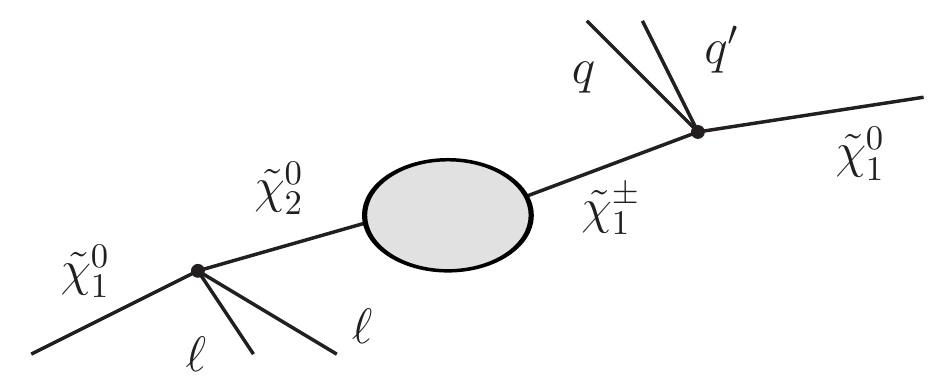}
\caption{Decay chain of interest for step \AGRSchaneuqqll{} (\agrs{} scenario).}
\label{fig:AGRSchaneu}
}
\subsubsection*{Step \AGRSchaneuqqll{}: \AGRSchaneuqqllTEXT}

\noindent Still to be determined is the $\chai$ mass. At variance with the \grs{}
scenario, this information cannot be easily extracted from secondary $\chai$
generated in $\go \go$ or $\stopi \stopi$ production events. However, one can
consider $\chai \neuii$ associated production, with the $\chai$ decaying as
$\chai \to \neui q q'$. Concerning the $\neuii$, the radiative decay to $\neui$ is,
as already noted above, the one with the largest branching ratio. However, the
resulting event topology, namely 2 $q$ + $\gamma$ + $\sla p_T$, is dominated by background
events, that we were not able to get rid of. For this reason, we required the $\neuii$
to decay as $\neui \ell \ell$ (and veto on $\ell = \tau$, since its decay would
produce missing momentum from neutrinos). The invariant-mass distribution of the
$q q'$ pair is bounded from above by $m_{\tilde \chi_1^\pm} - m_{\tilde \chi_1^0}$,
allowing to obtain $m_\chai$, since $m_\neui$ is known from step \AGRSgogo{}.

In addition, the invariant-mass distribution $m_{\ell \ell}$ provides a cross-check with
step \AGRSchaneulll{} on the determination of $m_\neuii - m_\neui$.

\section{Analysis and results} \label{sec:results}

\newcommand{\met}{$\sla E_T$}
\newcommand{\et}{$E_T$}
\newcommand{\mtgen}{$m_{T {\rm Gen}}$}

In this section we would like to show that the mass-determination
strategy described in the previous section is indeed practicable
with real data collected at the Atlas \cite{ATLAS-DTP} and CMS
\cite{CMS-TDR} experiments, and permits to determine most of the
considered masses within about 20 GeV of the true values, with an
integrated luminosity not exceeding 100 fb$^{-1}$ at the design LHC
center-of-mass energy of 14 TeV. To address this point, we have
simulated full events with Pythia 6.42 \cite{Pythia6.4} and modeled
LHC detector effects with PGS4 \cite{PGS4}. Specifically, we adopted
the default {\tt lhc.par} card that comes with the latest PGS4
version. Purpose of this analysis is to get a presumably realistic
idea of how much do the signals degrade once detector effects are
taken into account. It goes without saying that the `true' answer
will require a full-fledged detector simulation, with insider
knowledge of all acceptance and resolution details, which is
prerogative of experimentalists only. 

We next proceed to the discussion of the analysis, following the steps 
described in the previous section.

\subsection{Analysis: \grs{} scenario}

\subsubsection*{Step \GRSgogo{}: \GRSgogoTEXT}

\noindent We select events with 4 jets tagged as $b$, \footnote{For
the sake of clarity, the $b$-tagging algorithm adopted in our
analysis is the one defined as `loose tagging' in PGS \cite{PGS4}.
Needless to say, a fully rigorous treatment of this issue requires a
dedicated experimental analysis.} 4 $q$ jets (here and henceforth,
we will so indicate jets not tagged as $b$) and 2 leptons. To
reconstruct the $W$ bosons, we make all the possible di-jet
combinations out of the 4 $q$ jets, and accept only events for which
{\em both} di-jet invariant masses for at least one pairing are
within $M_W \pm 20$ GeV. In case that more than one paring
fulfill this requirement, we choose the pairing with minimal
$\sqrt{\Delta M_W^2(2q)+\Delta M_W^2(2q^\prime)}$. Furthermore, in 
order to eliminate possible SM backgrounds, we impose the following cuts:

\begin{itemize}
\item[{\bf (a)}] $p_{T 1,2,3,4} >$ 50, 30, 20, 20 GeV for the 4 $b$ jets;
\item[{\bf (b)}] Missing transverse energy \met $>$ 50 GeV;
\item[{\bf (c)}] Transverse sphericity $S_T >$ 0.15.
\footnote{We adopt the same definition as \cite{ATLAS-DTP}, namely
\beq
S_T ~\equiv~ \frac{2 \lambda_2}{\lambda_1 + \lambda_2}~,
\eeq
where $\lambda_{1,2}$ are the eigenvalues of the $2 \times 2$ sphericity tensor
$S_{ij} = \sum_k p^k_i p^k_j$, with $k$ running on all the visible particles, and $i,j =
\{ 1,2 \}$ labeling the momentum transverse component. SUSY events tend to be spherical ($S_T \sim 1$)
as opposed to back-to-back.}
\end{itemize}

As this process involves several jets, in order to reduce the uncertainty
associated with jet momentum resolution we use here the $\mtt$
defined in terms of the total transverse mass, $m_{TV}$, of the 4 jets in
the gluino decay $\tilde{g}\rightarrow t\tilde{t}_1\rightarrow
bqqb\tilde{\chi}_1^{\pm}$, see eq. (\ref{MTdef2}). Furthermore, in the procedure 
to compute $m_{TV}$, we do not use the invariant mass of each jet
recorded in the detector, but simply set $m_j=0$ for non-$b$ jets,
and $m_j=m_b$ for $b$-tagged jets. This means that, for a given trial
chargino mass, only the jet transverse momenta are used for the
calculation of $\mtt$. Quite often, this approach turns out to be
useful for identifying the endpoint when the decay product of each
mother particle involves more than one (or two) jets. Still, the
calculation of $\mtt$ is subject to combinatoric uncertainty due to
correctly pairing the 4 $b$ jets with the two reconstructed $W$
bosons. To perform this pairing, we use the so-called \mtgen{} pairing 
scheme \cite{MTGEN}.%
\footnote{In the present example, the \mtgen{} scheme works as follows. Assuming that only one
of the (three) di-jet pairings among $q$ jets fulfills the $M_W$ requirement, there are still six
possible ways of forming two $b$ di-jets and of assigning them to the two reconstructed $W$ bosons.
For each of these combinations, we calculate the $\mtt(m_\chi)$ value for fixed $m_\chi$, and pick
the combination corresponding to the minimum $\mtt(m_\chi)$. As $m_\chi$ value we took 100 GeV and
we checked that the pairing choice does not change when reasonably varying $m_\chi$.}
\begin{figure}[ht]
\begin{center}
\includegraphics[width=0.45 \textwidth]{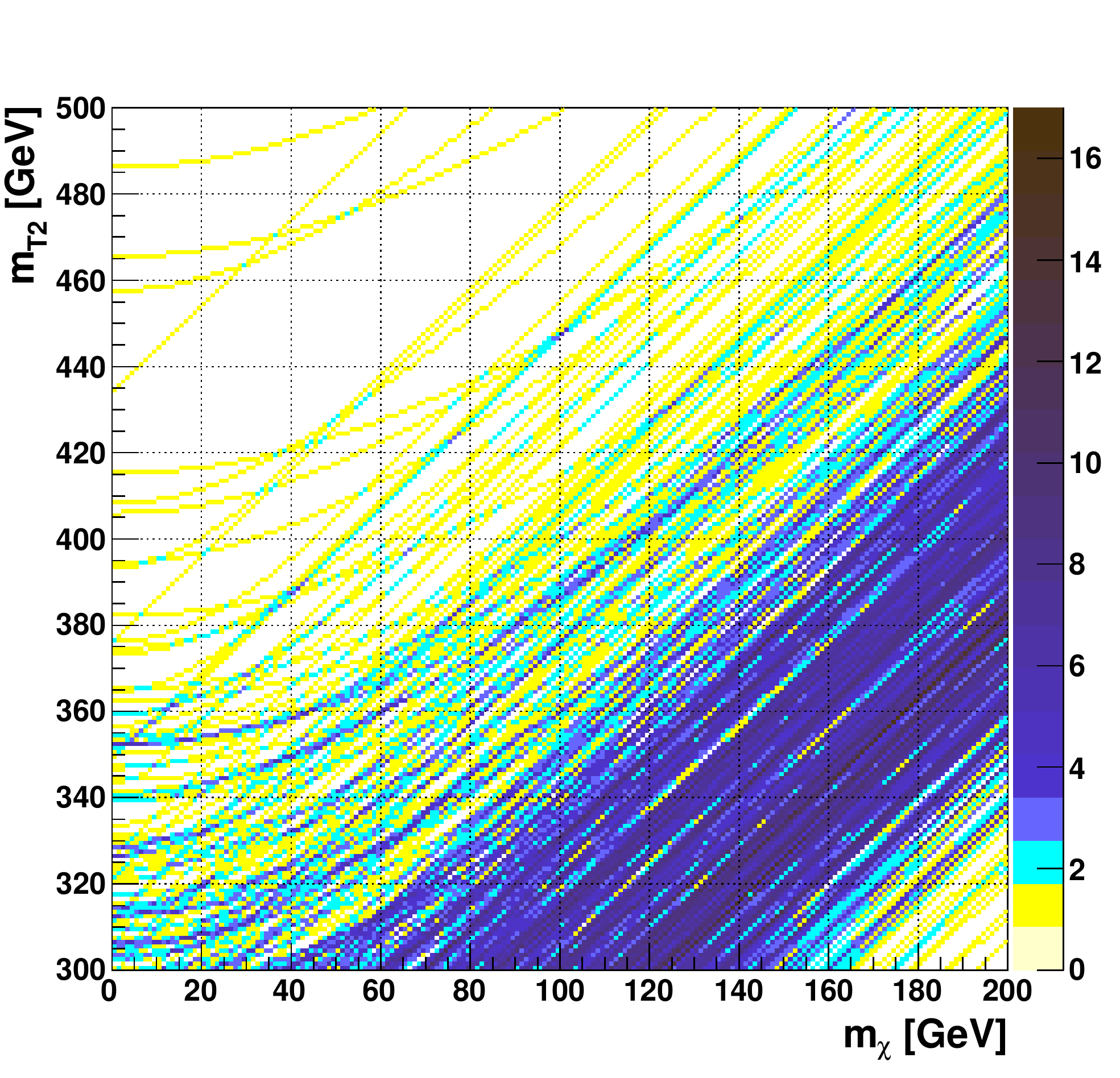} \hfill
\includegraphics[width=0.45 \textwidth]{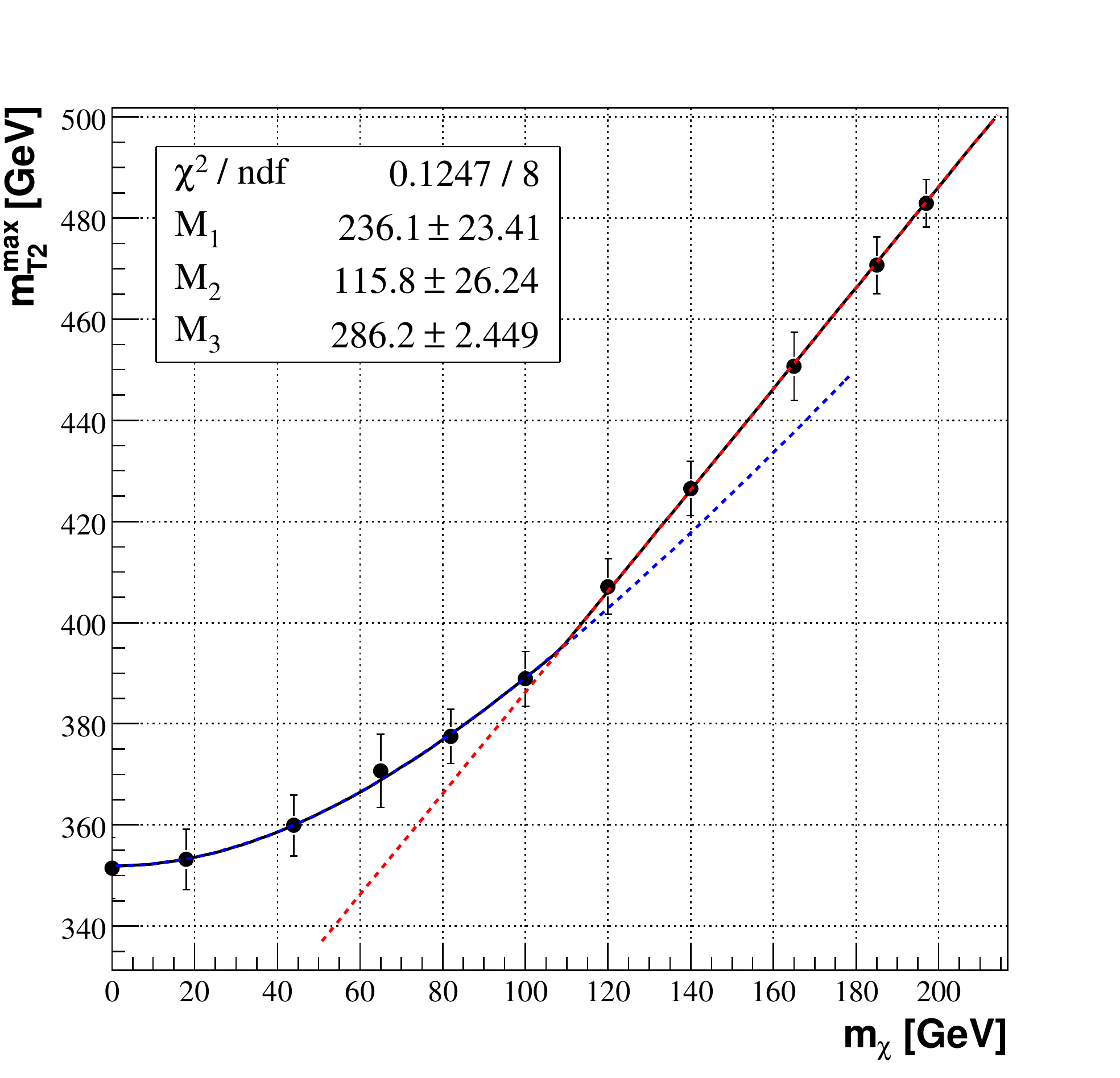}
\vspace{1cm}
\includegraphics[width=0.32 \textwidth]{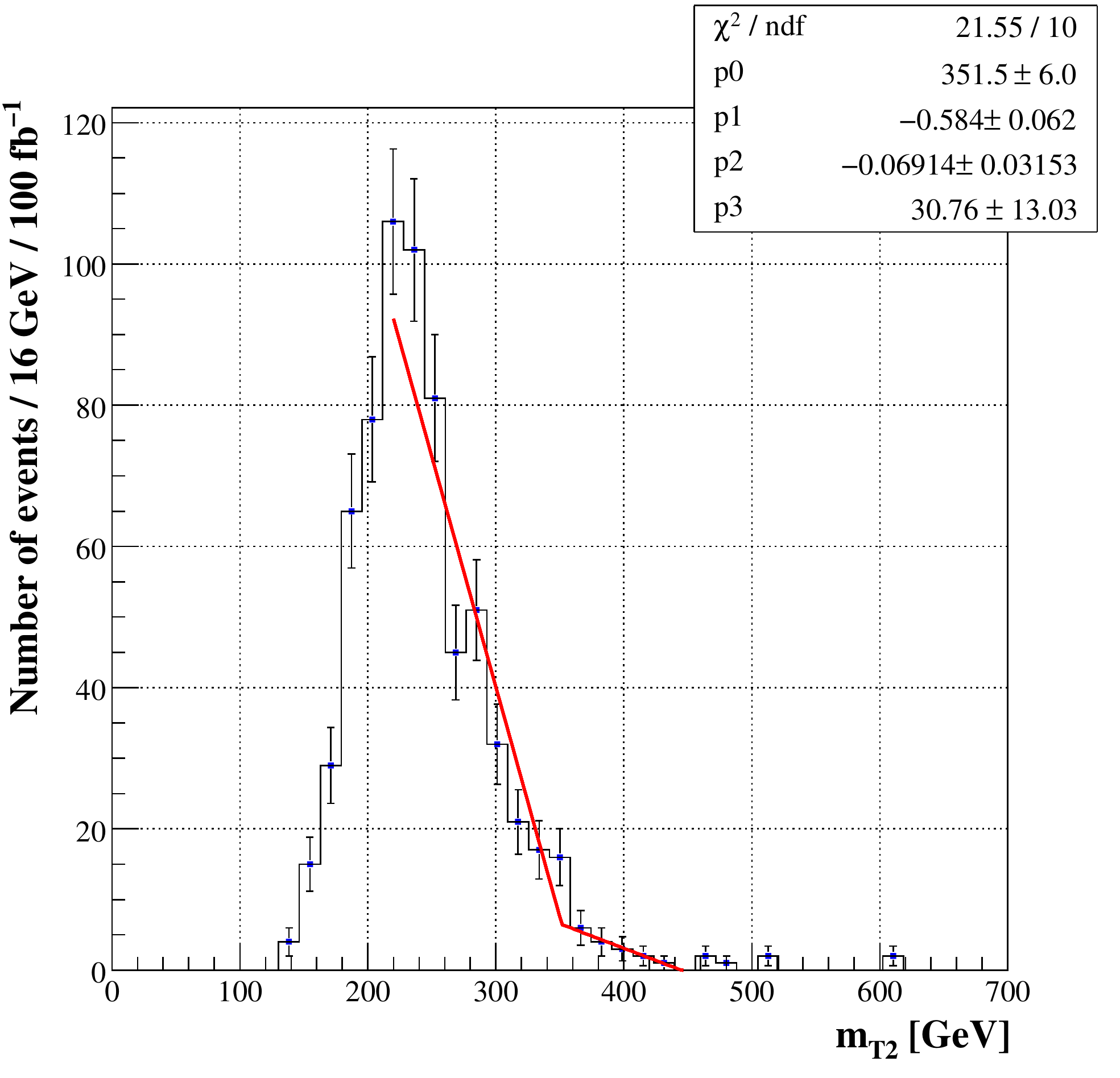}
\includegraphics[width=0.32 \textwidth]{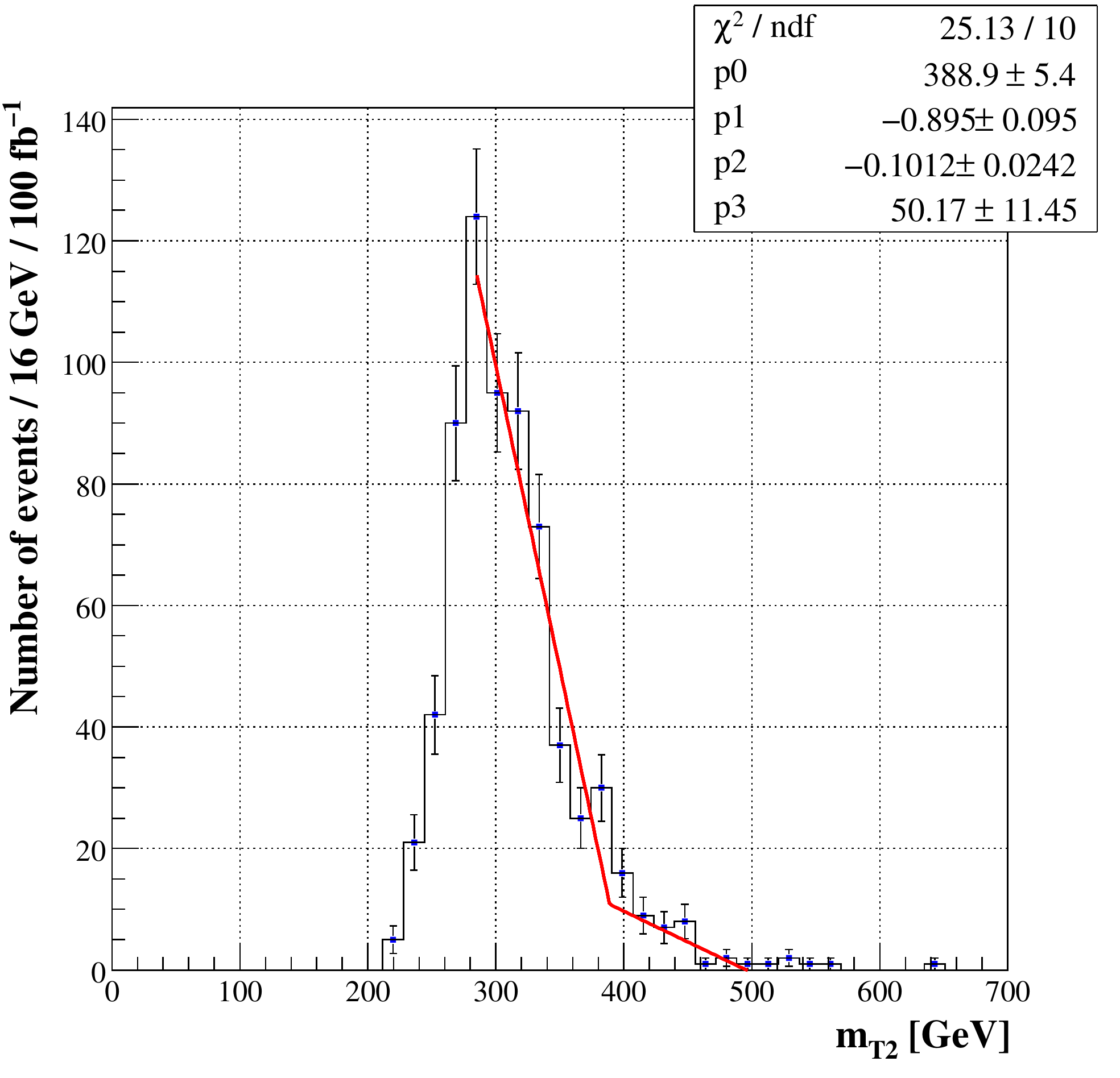}
\includegraphics[width=0.32 \textwidth]{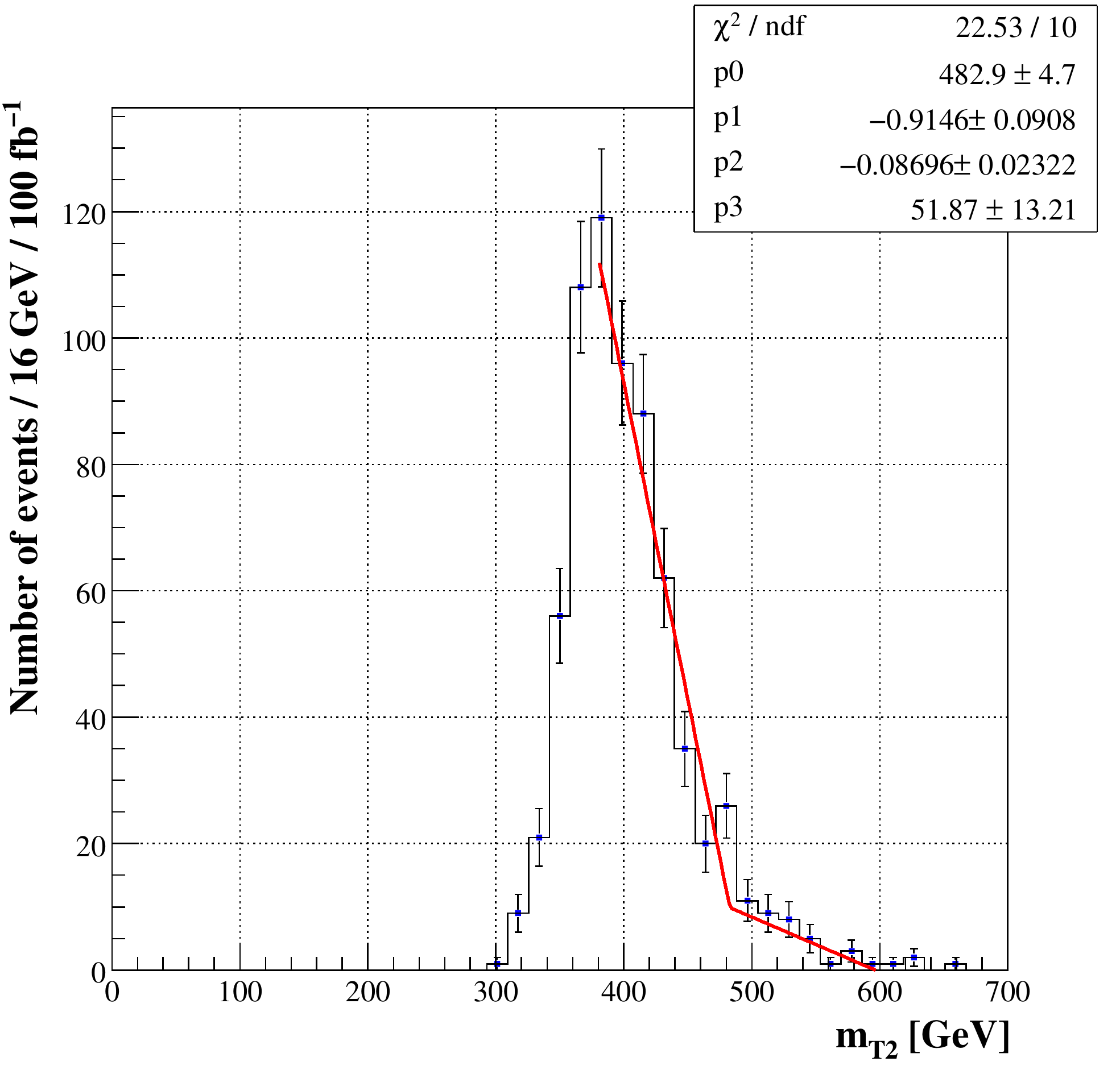}
\caption{{\em Upper-left panel:} $\mtt(m_\chi)$ density plot for the events of step \GRSgogo{}, \grs{} scenario.
The color code on the right of the plot represents the number of events described by each line;
{\em Upper-right panel:}  fit to the corresponding maximum $\mtt(m_\chi)$;
{\em Lower panels:} $\mtt$ distribution for the events of step \GRSgogo{}, \grs{} scenario, with
$m_\chi = \{0, 100, 197\}$ GeV.}
\label{fig:GRSgg-mT2}
\end{center}
\end{figure}

With the events passing the above selection criteria,%
\footnote{In particular, we checked by a parton-level analysis that
the adopted \mtgen{} pairing scheme is effective in reducing the jet
combinatoric error, while not significantly affecting the final
$\mtt$ distribution with respect to the true pairing.} we have
constructed the $\mtt(m_\chi)$ distribution, with trial 
chargino mass $m_\chi \in [ 0, 200]$ GeV. The resulting density
plot is shown in fig. \ref{fig:GRSgg-mT2} (upper-left panel). 
The kink structure can be roughly located by inspecting the uppermost 
dark lines (indicating the largest density of events), if one neglects
the more sparse (lighter in color) contributions from background
events.

A quantitative determination of the $\mtt$ maximum line can be
carried out as follows. First, one fits the endpoint of the $\mtt$
distributions obtained at fixed $m_\chi$. Examples of these
distributions, and their endpoint fits to a two-segments line, are
reported in fig. \ref{fig:GRSgg-mT2} (lower panels) for the values
$m_\chi = \{ 0, 100, 197 \}$ GeV. Here and henceforth (figs. 
\ref{fig:GRSgg-mT2}-\ref{fig:AGRS-steps23}), the two-segments line is parameterized 
as
\beq
\label{eq:twosegments}
\begin{tabular}{ll}
$y ~=~ p_1 (x - p_0) + p_2 x + p_3~,$ & $~~~\mbox{for }x < p_0~,$\\
$y ~=~ p_2 x + p_3~,$ & $~~~\mbox{for }x > p_0~.$ \\
\end{tabular}
\eeq
The resulting endpoint values 
of $\mtt$ for the various values of $m_\chi$, namely $\mtt^{\rm max}
(m_\chi)$, are shown in fig. \ref{fig:GRSgg-mT2} (upper-right panel), 
with bars
representing the statistical error. To identify the kink position,
we then fit those endpoint values with appropriate fitting
functions. To this end, one can use the known analytic form of
$\mtt(m_\chi)$ \cite{kink1,kink2} for events with
$\mtt(m_\chi=m_\chai)=m_\go$ and negligible upstream transverse
momentum: 
\beqn \label{eq:fitfun} 
\mtt(m_\chi,m_V) = \frac{m_\go^2 - m_\chai^2+m_V^2}{2m_\go} + 
\frac{\sqrt{\left( m_\go^2 + m_\chai^2-m_V^2 \right)^2 + 4
m_\go^2\left(m_\chi^2-m_\chai^2\right)}}{2m_\go}~,
\eeqn 
where $m_V$
denotes the total invariant (or transverse) mass of the $2b + 2q$ jets in
the gluino decay. From this $\mtt$ functional form, one can easily notice
that $\mtt^{\rm max}=\mtt(m_V=m_V^{\rm max})$ for $m_\chi>m_\chai$,
while $\mtt^{\rm max}=\mtt(m_V=m_V^{\rm min})$ for $m_\chi<m_\chai$,
hence $\mtt^{\rm max}$ shows a kink at $m_\chi=m_\chai$.
Generically, $m_V^{\rm max}\leq m_\go-m_\chai$ and $m_V^{\rm min}\geq 0$, 
and one easily finds $\mtt^{\rm max}=m_\go-m_\chai+m_\chi$ for
$m_V^{\rm max}=m_\go-m_\chai$.

The $\mtt$ endpoint values in fig. \ref{fig:GRSgg-mT2} for
$m_\chi > m_\chai$ indicate that they can be well described by
a straight line, implying that $m_V^{\rm max}$ is close to
$m_\go - m_\chai$. Inspired by the analytic form (\ref{eq:fitfun}), we
 then use the following fitting functions to find the kink position:
\beq
\label{fitfun1}
\begin{tabular}{ll}
$\mtt^{\rm max} = M_1 +\sqrt{M_2^2+m_\chi^2}$ & ~~~if ~$m_\chi < m_\chai$~,\\
$\mtt^{\rm max} = M_3+m_\chi$ & ~~~if ~$m_\chi > m_\chai$~,
\end{tabular}
\eeq
where $M_i$ ($i=1,2,3$) are fitting parameters.
The best-fit curves are shown in fig. \ref{fig:GRSgg-mT2}
(upper-right panel), together with the corresponding values of
$M_i$. The resulting gluino and chargino masses are given by
$m_\chai = 109(17)$ GeV and $m_\go = 395(16)$ GeV, which are fully compatible
with the true values in table \ref{tab:examplefits}. We also find
that the $\mtt^{\rm max}$-curves in fig. \ref{fig:GRSgg-mT2}, which
describe data at detector level, are reasonably close to the curves 
obtained at parton level.


\subsubsection*{Step \GRStt{}: \GRSttTEXT}

To select these events, we need to require 2 jets tagged as $b$, 4 $q$ jets, and no leptons.
The channel of $\stopi$ pair production is subject to SM backgrounds coming mostly from 
$t \bar t$ or $W W + 2 b$ production. These backgrounds can be eliminated efficiently 
by the following selection cuts
\begin{itemize}
\item[{\bf (a)}] $p_{T 1,2} >$ 50, 25 GeV on the 2 $b$ jets;
\item[{\bf (b)}] $p_{T 1,2,3,4} >$ 50, 25, 20, 10 GeV on the 4 $q$ jets;
\item[{\bf (c)}] \met $>$ 100 GeV;
\item[{\bf (d)}] Transverse sphericity $S_T >$ 0.15.
\end{itemize}

\begin{figure}[ht]
\begin{center}
\includegraphics[width=0.40 \textwidth]{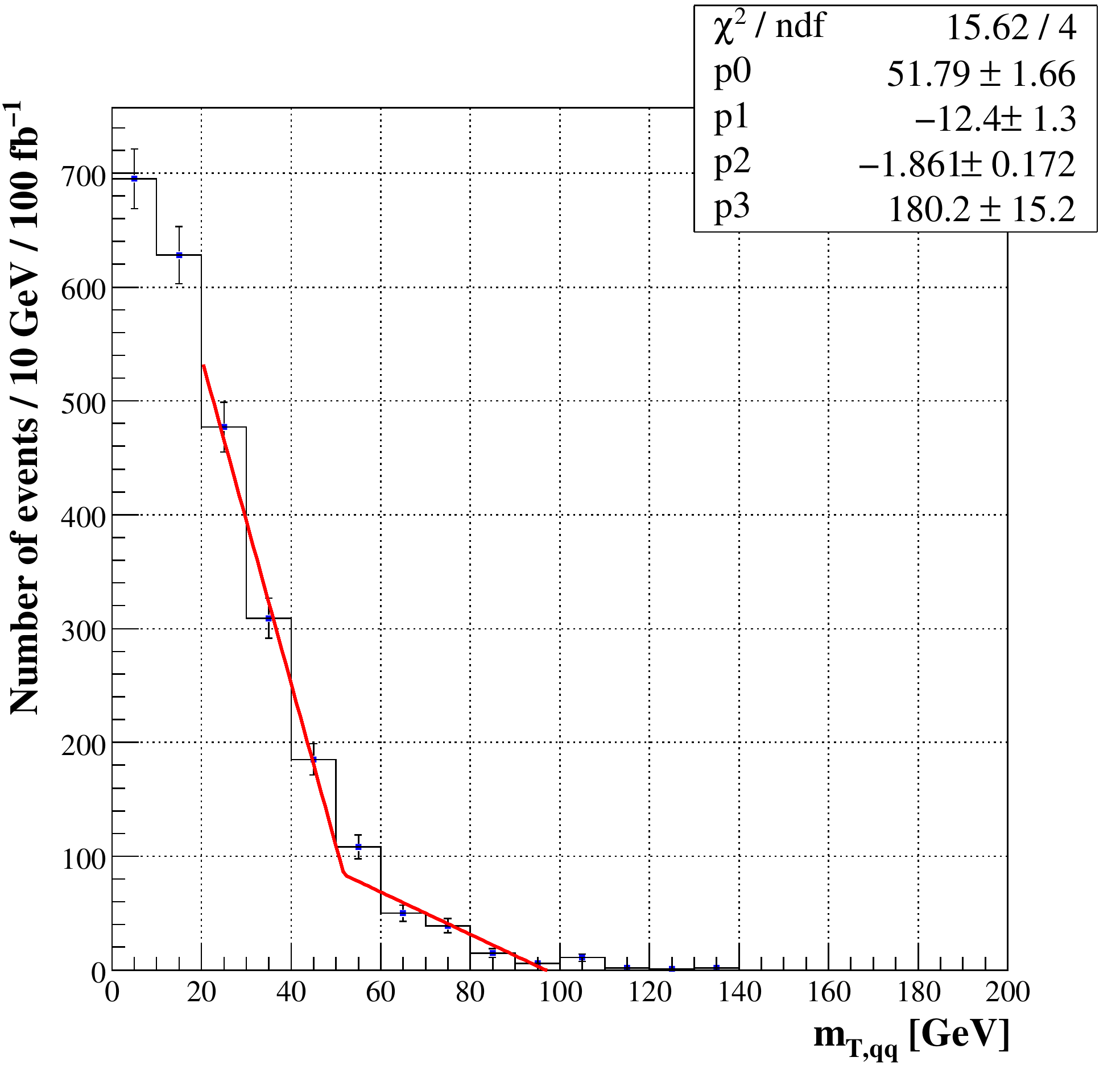} \hfill
\includegraphics[width=0.40 \textwidth]{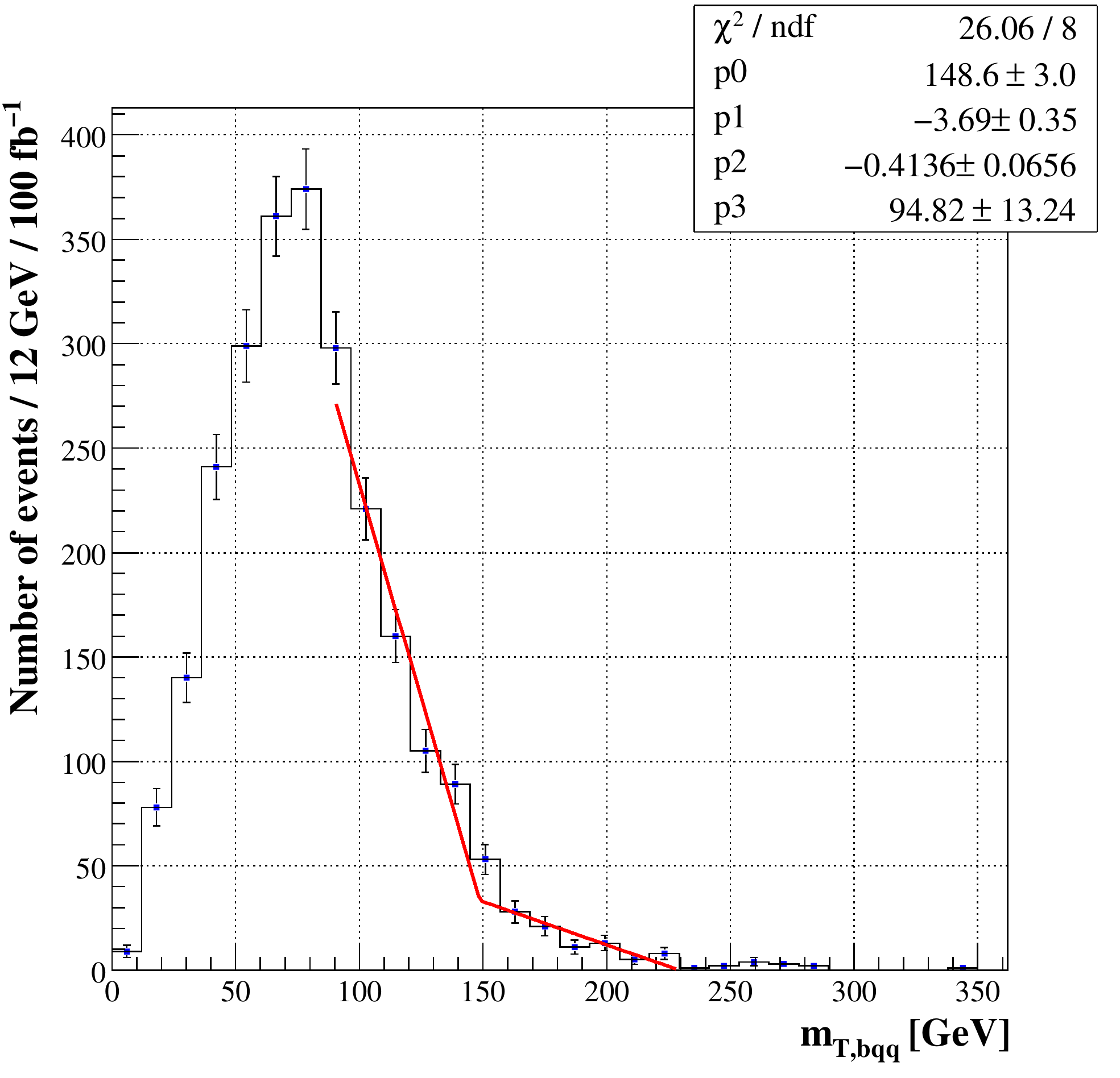} \\
\includegraphics[width=0.40 \textwidth]{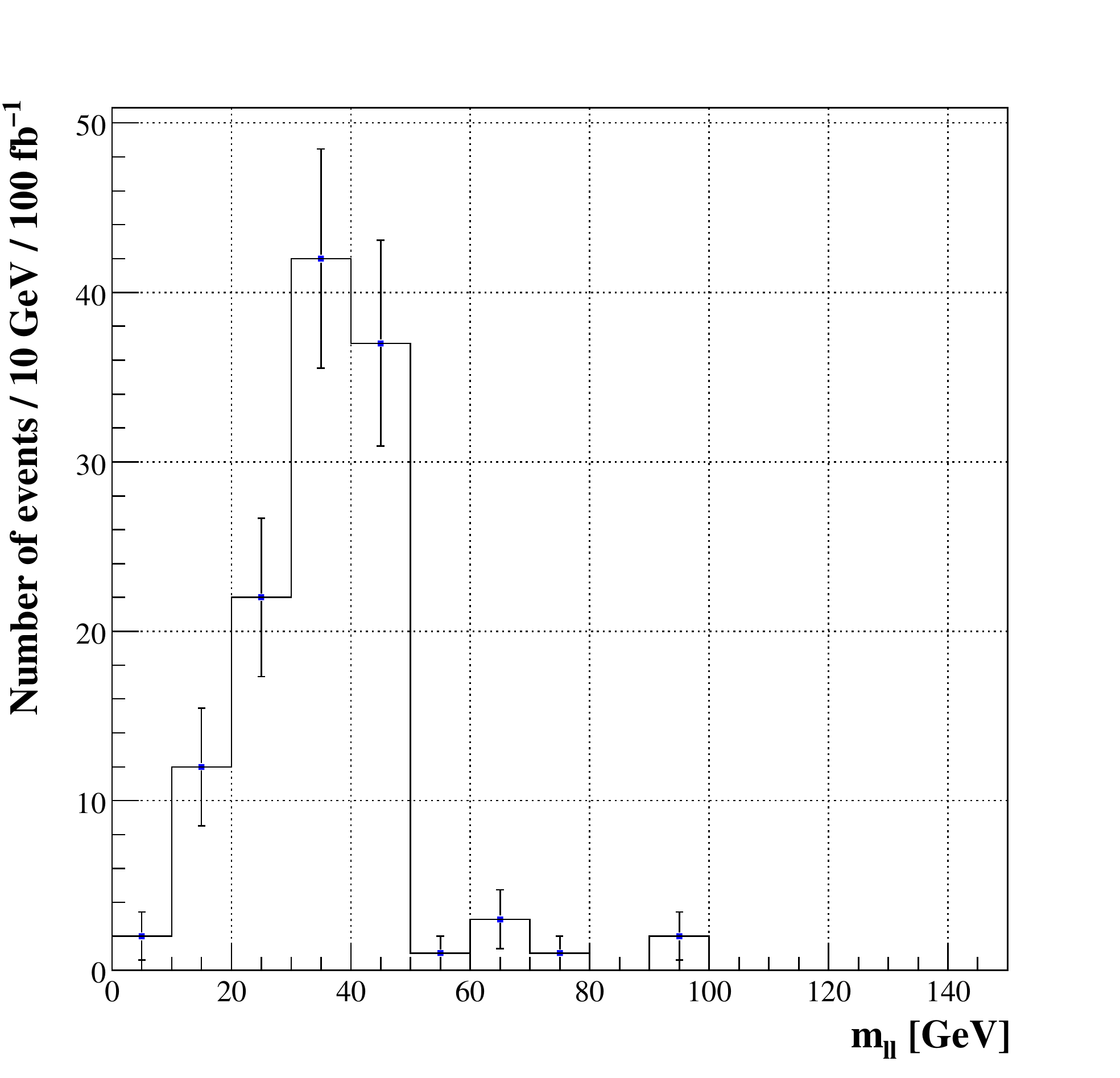}
\caption{{\em Upper-left panel:} transverse mass distribution $m_{T,qq}$ (step \GRStt{}, \grs{} scenario);
{\em Upper-right panel:} $bqq$-jet transverse mass distribution (step \GRStt{}, \grs{} scenario);
{\em Lower panel:} invariant mass distribution of the $\ell^+ \ell^-$ pair (step \GRSchaneulll{}, \grs{} scenario).}
\label{fig:GRS-steps23}
\end{center}
\end{figure}

For the events passing the above cuts (a)-(d), we calculated, event
by event, all possible di-jet transverse mass values out of the 4
$q$ jets and picked the minimum value among them. Here we use
the di-jet transverse mass $m_{T,jj}$, rather than the di-jet
invariant mass $m_{jj}$, as the endpoint of $m_{jj}$ is severely
smeared. The corresponding distribution is shown in fig.
\ref{fig:GRS-steps23} (upper-left panel). Fitting the endpoint
region to a two-segments line, we obtain $m_{T,jj}^{\rm max}=m_\chai
- m_\neui = 51.8(1.7)$ GeV, corresponding to $m_\neui \approx 58$
GeV, a good estimate of the true value in table
\ref{tab:examplefits}. A completely analogous procedure can be
adopted for the transverse mass of the $b q q'$ system, and the
resulting distribution is shown in fig. \ref{fig:GRS-steps23}
(upper-right panel). The fitted endpoint $m_{T,bqq}^{\rm max}$
of this distribution allows to determine the $\stopi$ mass through
the following formula 
\beq 
m_\stopi ~=~ \sqrt{\frac{(m_{T,bqq}^{\rm max})^2+m_\chai^2-m_\neui^2}
{1-m_\neui^2 / m_\chai^2}}~,
\label{eq:mstop} 
\eeq
implying $m_\stopi \approx 206$ GeV, again consistent with the true 
value in table \ref{tab:examplefits}.%
\footnote{Note that, if the relation $m_\chai^2 = m_\stopi m_\neui$
is fulfilled (as is our case within about (25 GeV)$^2$), eq.
(\ref{eq:mstop}) simplifies to $m_{T,bqq}^{\rm max} = m_\stopi - m_\neui$.}

\subsubsection*{Step \GRSchaneulll{}: \GRSchaneulllTEXT}

\noindent To select these events, we require 2 leptons (either $e$ or $\mu$) of same flavor and opposite
charge, one additional lepton of different flavor (with veto on hadronically decaying taus), and no jets.
To clean up the event selection from possible backgrounds, we also impose the following cuts:

\begin{itemize}
\item[{\bf (a)}] $p_{T 1,2} >$ 20, 15 GeV on the 2 leptons of same flavor;
\item[{\bf (b)}] $p_{T} >$ 15 GeV on the lepton of different flavor;
\item[{\bf (c)}] \met $>$ 50 GeV;
\item[{\bf (d)}] Transverse sphericity $S_T >$ 0.15.
\end{itemize}
The invariant mass distribution constructed with the 2 leptons of same flavor is shown in fig.
\ref{fig:GRS-steps23} (lower panel). The fact that, in general, the lepton signal is quite accurately
reconstructed explains the clear end point structure of the invariant mass distribution at 50 GeV (even
with a limited number of events). This endpoint value is in agreement with the expected
$m_\neuii - m_\neui \doteq 51$ GeV from table \ref{tab:examplefits}.

\subsection{Analysis: \agrs{} scenario}

\subsubsection*{Step \AGRSgogo{}: \AGRSgogoTEXT}

\begin{figure}[ht]
\begin{center}
\includegraphics[width=0.40 \textwidth]{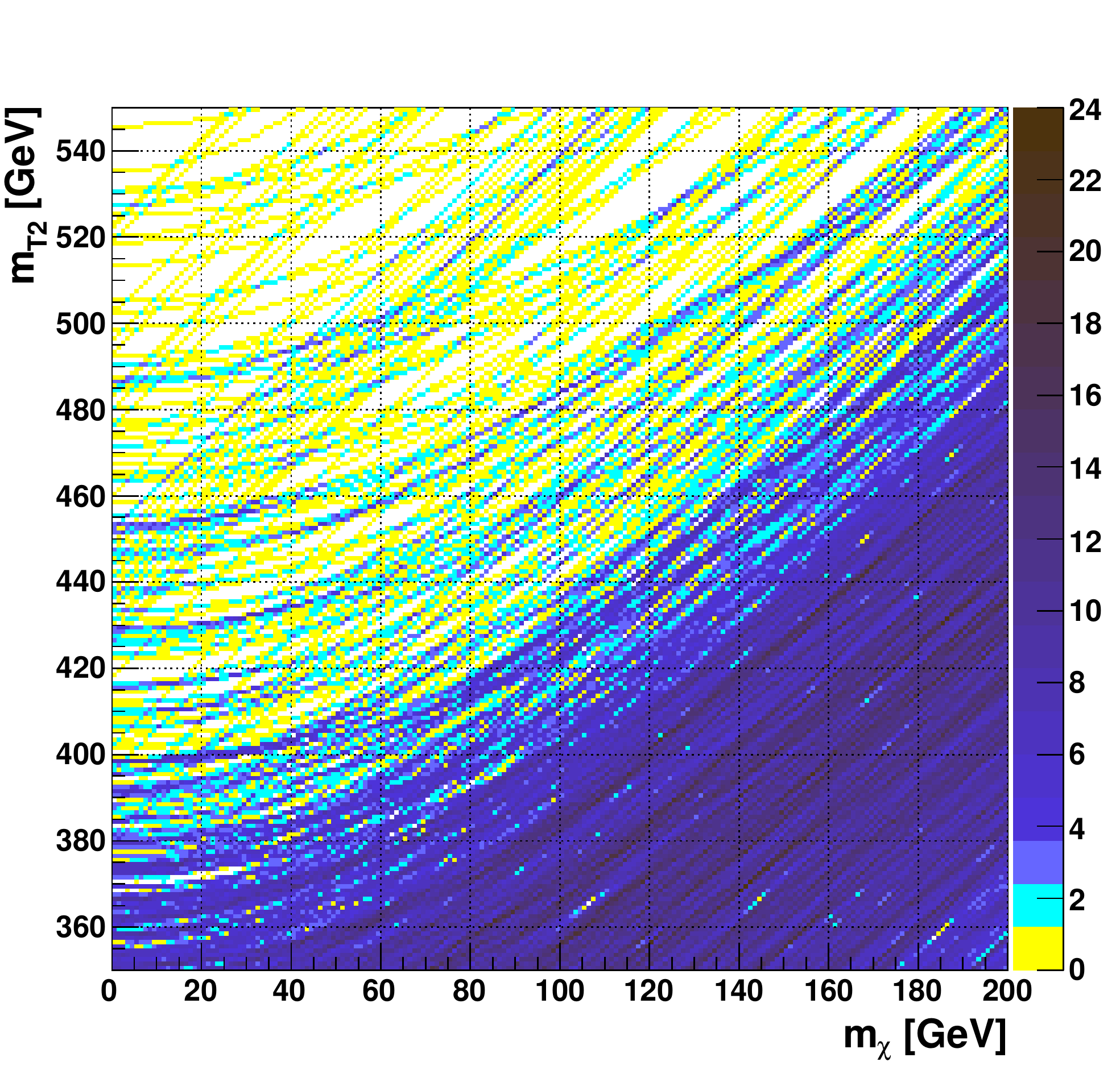} \hfill
\includegraphics[width=0.40 \textwidth]{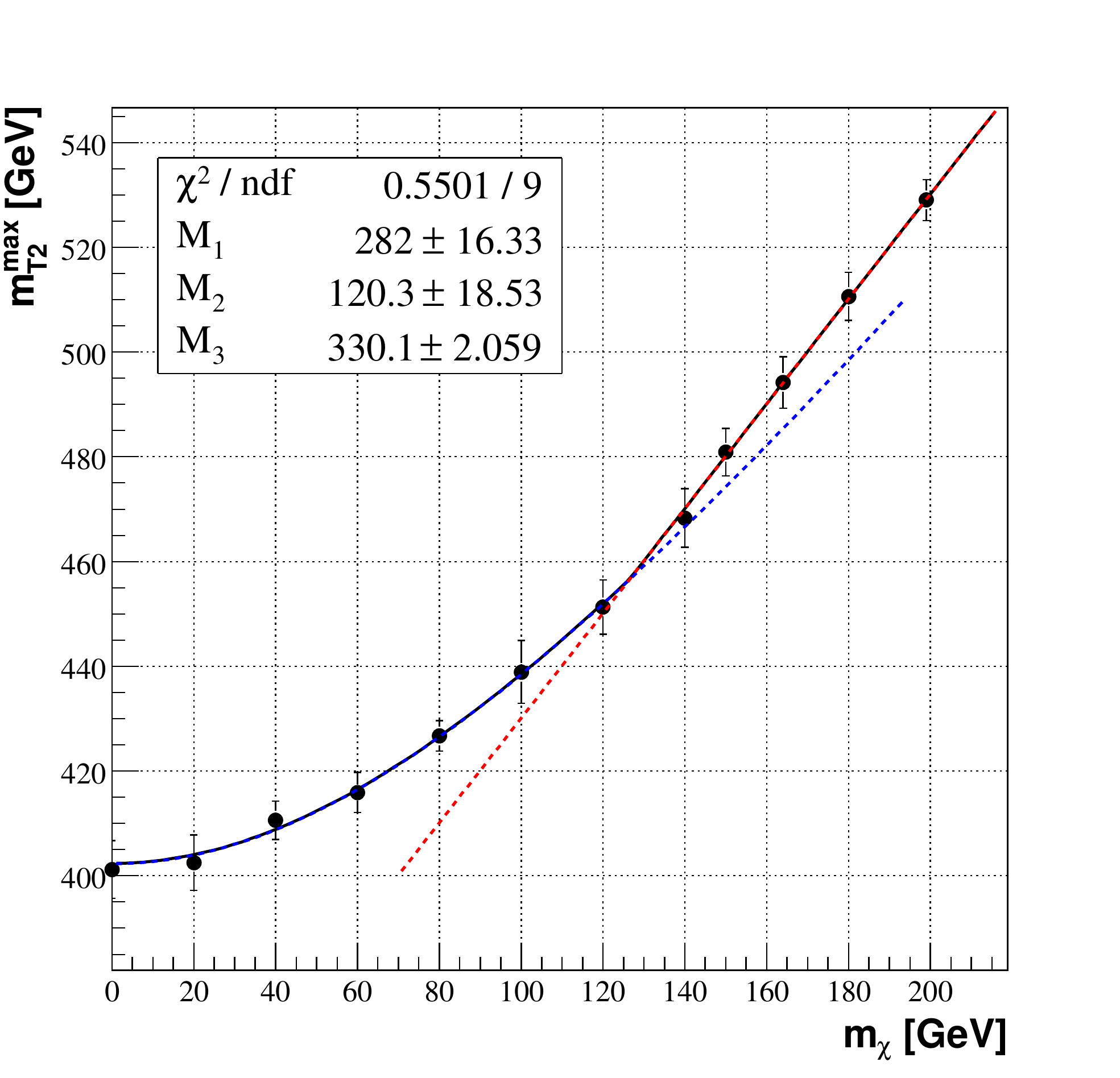}
\vspace{1cm}
\includegraphics[width=0.30 \textwidth]{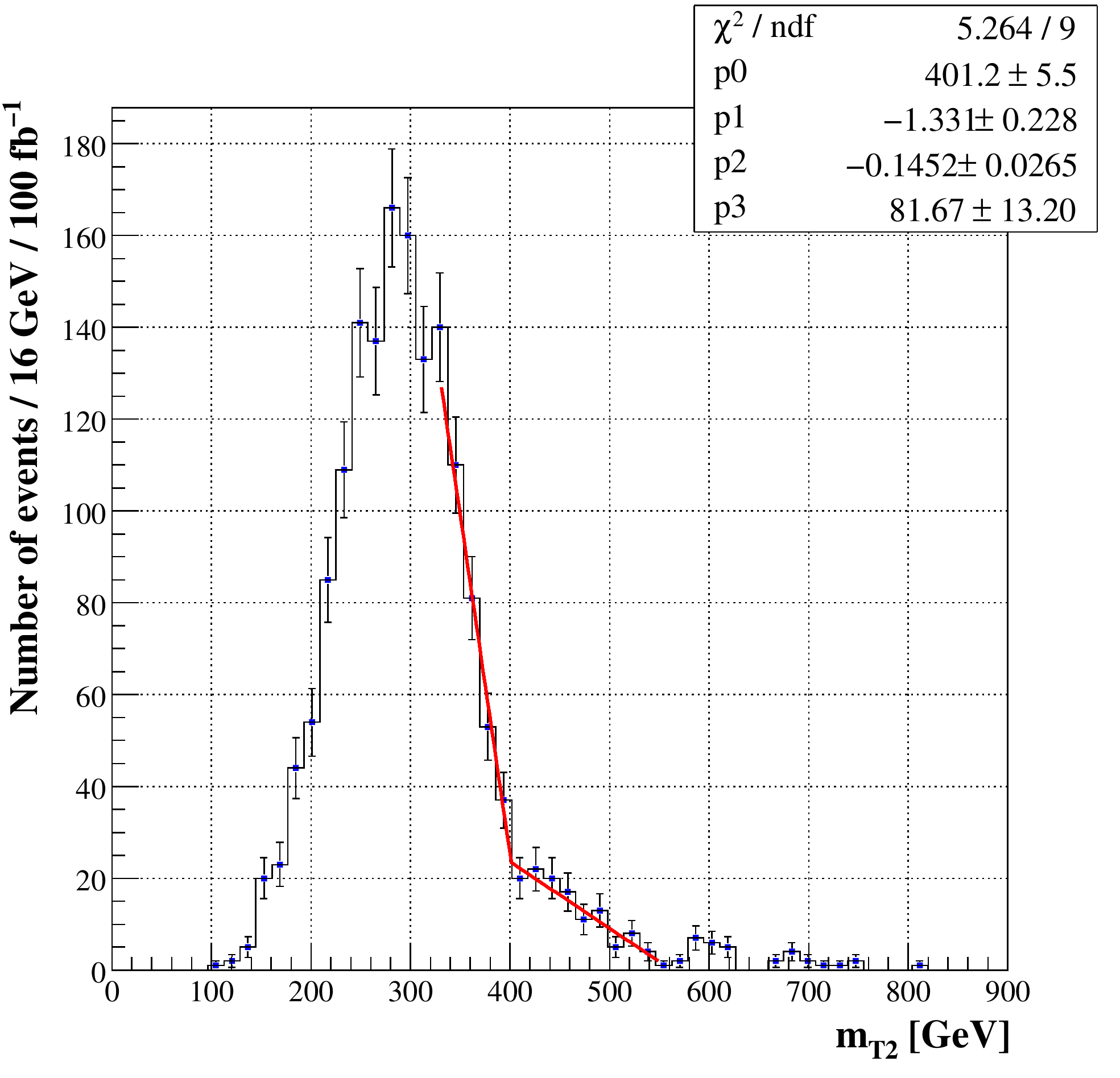}
\includegraphics[width=0.30 \textwidth]{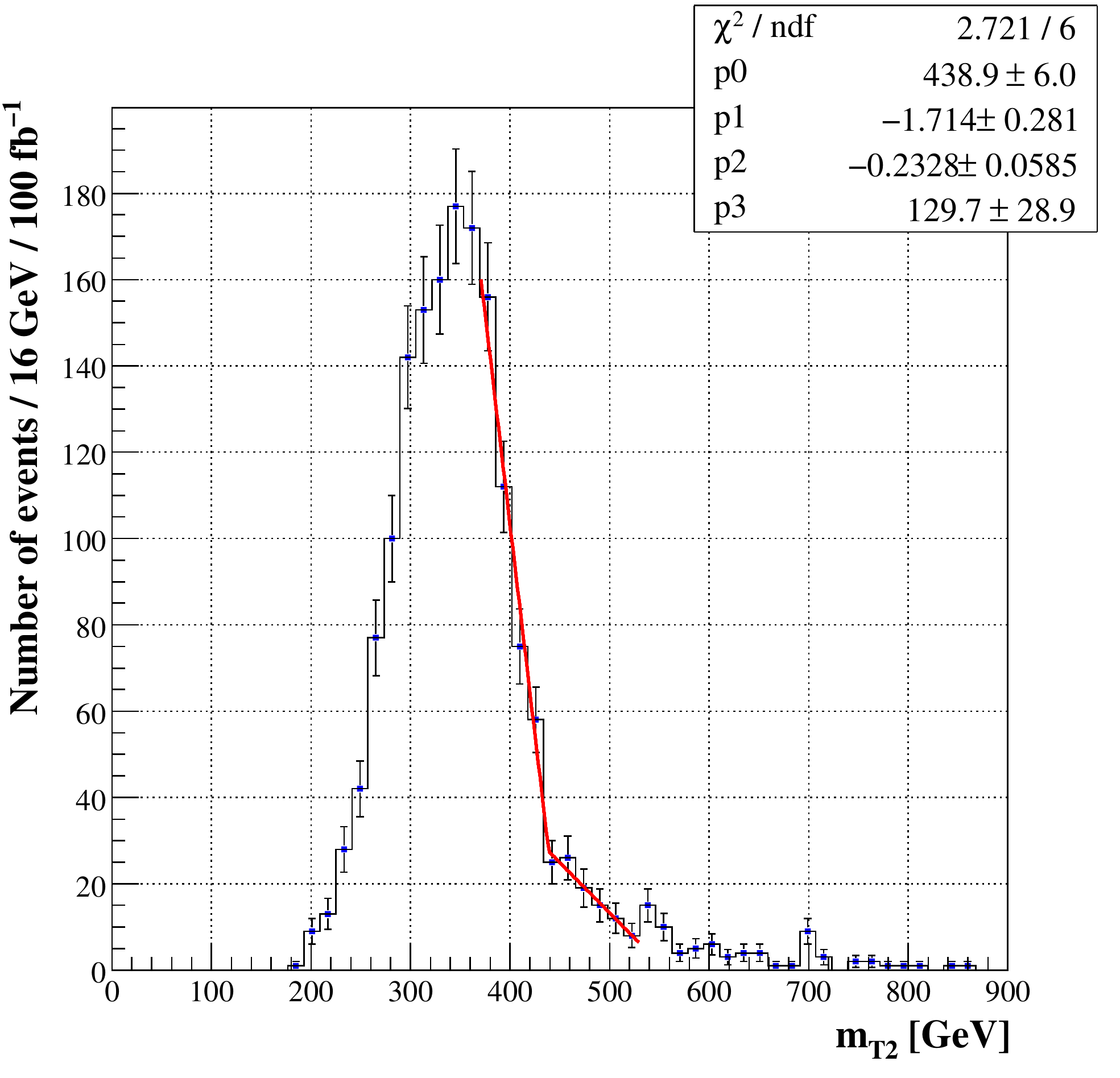}
\includegraphics[width=0.30 \textwidth]{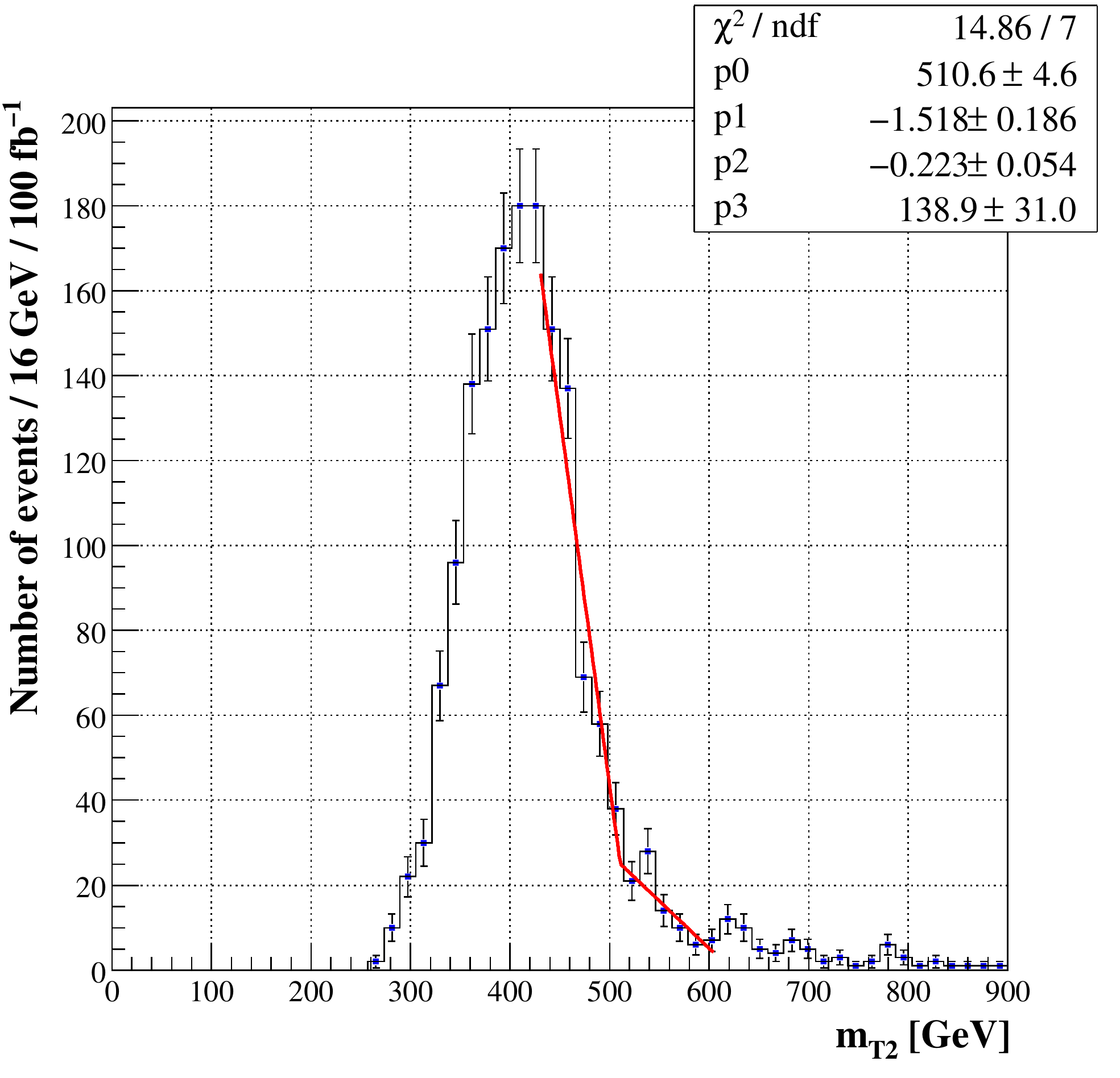}
\caption{{\em Upper-left panel:} $\mtt(m_\chi)$ density plot for the events of step \AGRSgogo{}, \agrs{} scenario.
The color code on the right of the plot represents the number of events described by each line;
{\em Upper-right panel:}  fit to the corresponding maximum $\mtt(m_\chi)$;
{\em Lower panels:} $\mtt$ distribution for the events of step \AGRSgogo{}, \agrs{} scenario, with
$m_\chi = \{0, 100, 180\}$ GeV.}
\label{fig:AGRSgg-mT2}
\end{center}
\end{figure}

\noindent Within $\go \go$ production, the decay $\go \to \neuii b \ov b$, followed by $\neuii \to \neui \gamma$
is the dominant channel. Since this channel is the only possible one giving 4 jets, 2 $\gamma$ and no lepton,
it may not be necessary to require $b$ tagging. Indeed, we select the above event content without $b$ tagging.
We however impose the following cuts as well:

\begin{itemize}
\item[{\bf (a)}] $p_{T 1,2,3,4} >$ 120, 70, 30, 20 GeV on the 4 jets;
\item[{\bf (b)}] $p_{T 1,2} >$ 50, 20 GeV on the hard photons;
\item[{\bf (c)}] \met $>$ 100 GeV;
\item[{\bf (d)}] Transverse sphericity $S_T >$ 0.15.
\end{itemize}

On the selected events, we calculate $\mtt(m_\chi)$ with the 4 jets
according to the \mtgen{} pairing scheme, and treating the 2
$\gamma$ as part of the \met. Here we use the di-jet invariant
mass $m_{jj}$, rather than the transverse mass, to compute $\mtt$, 
as the resulting $\mtt$ distribution shows a reasonably clean endpoint 
structure.

The $\mtt(m_\chi)$ density plot, displayed in fig.
\ref{fig:AGRSgg-mT2} (upper-left panel), shows a substantial number
of spurious events contributing to the maximum $\mtt$ line region,
impairing a `by eye' determination of the kink. Still, we can
quantify the kink position by following the same strategy as step
\GRSgogo{}, \grs{} scenario, namely fitting first the endpoint of
the $\mtt$ distributions obtained at fixed $m_\chi$ values. Fig.
\ref{fig:AGRSgg-mT2} (lower panels) shows those distributions and
fitted endpoints for $m_\chi = \{0, 100, 180\}$. Fitting those
endpoints (obtained for various values of $m_\chi$) to the function
in eq. (\ref{fitfun1}) (see fig. \ref{fig:AGRSgg-mT2}, upper-right
panel), we get the kink position at $m_\neuii = 126(16)$ GeV and $m_\go
= 456(15)$ GeV, which are in agreement with the true values in table
\ref{tab:examplefits}. (The best-fit values of $M_i$ are also
reported in the figure.) Comparing to the parton level result, we find
the kink to be not as sharp as expected. The main reason for this is
that a large portion of the events with small $m_{jj}$, that provide
the true $\mtt^{\rm max}$ for $m_\chi< m_\neuii$, are eliminated by
the selection cuts.

\subsubsection*{Step \AGRSchaneulll{}: \AGRSchaneulllTEXT}

\begin{figure}[ht]
\begin{center}
\includegraphics[width=0.40 \textwidth]{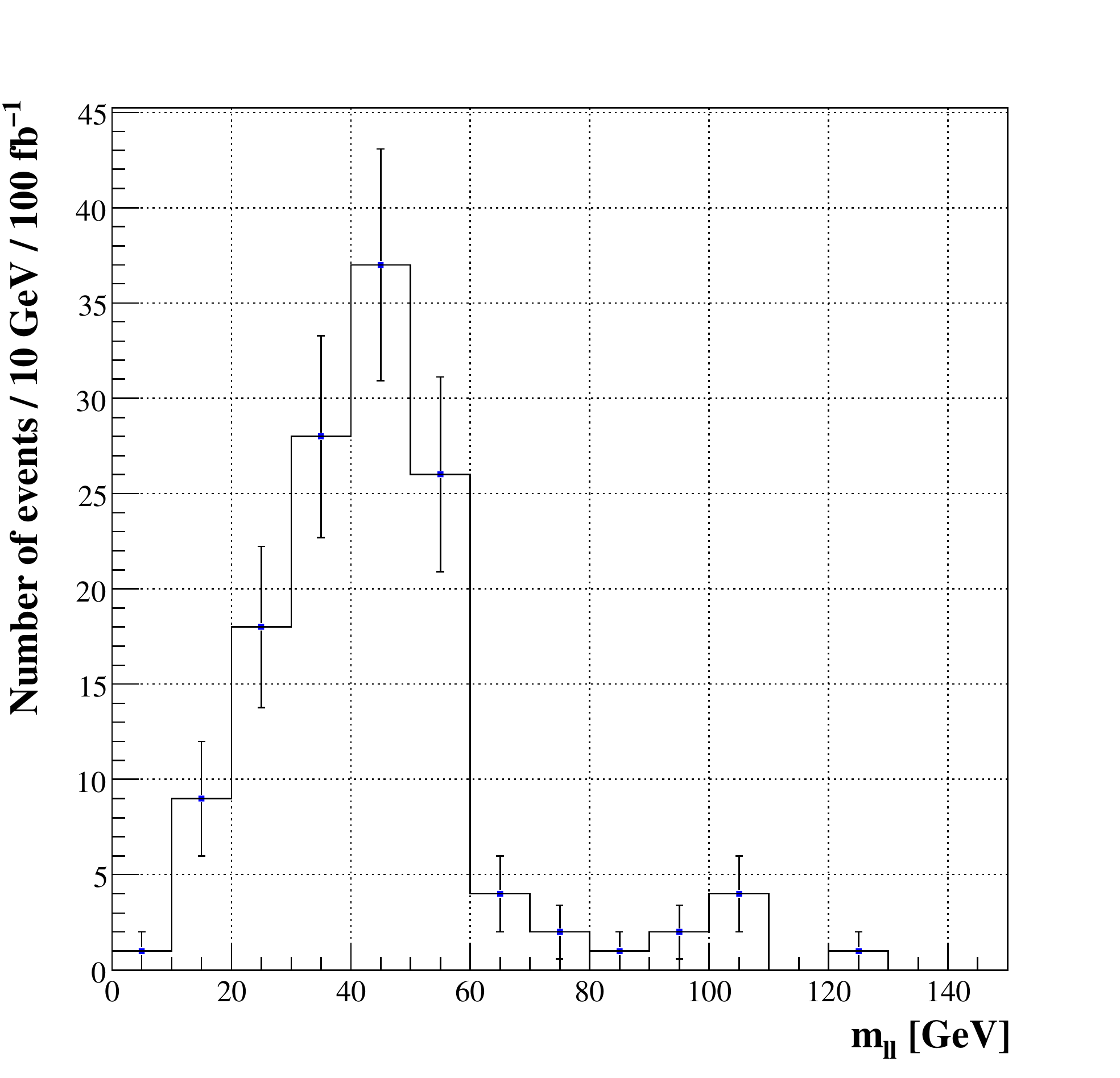} \hfill
\includegraphics[width=0.40 \textwidth]{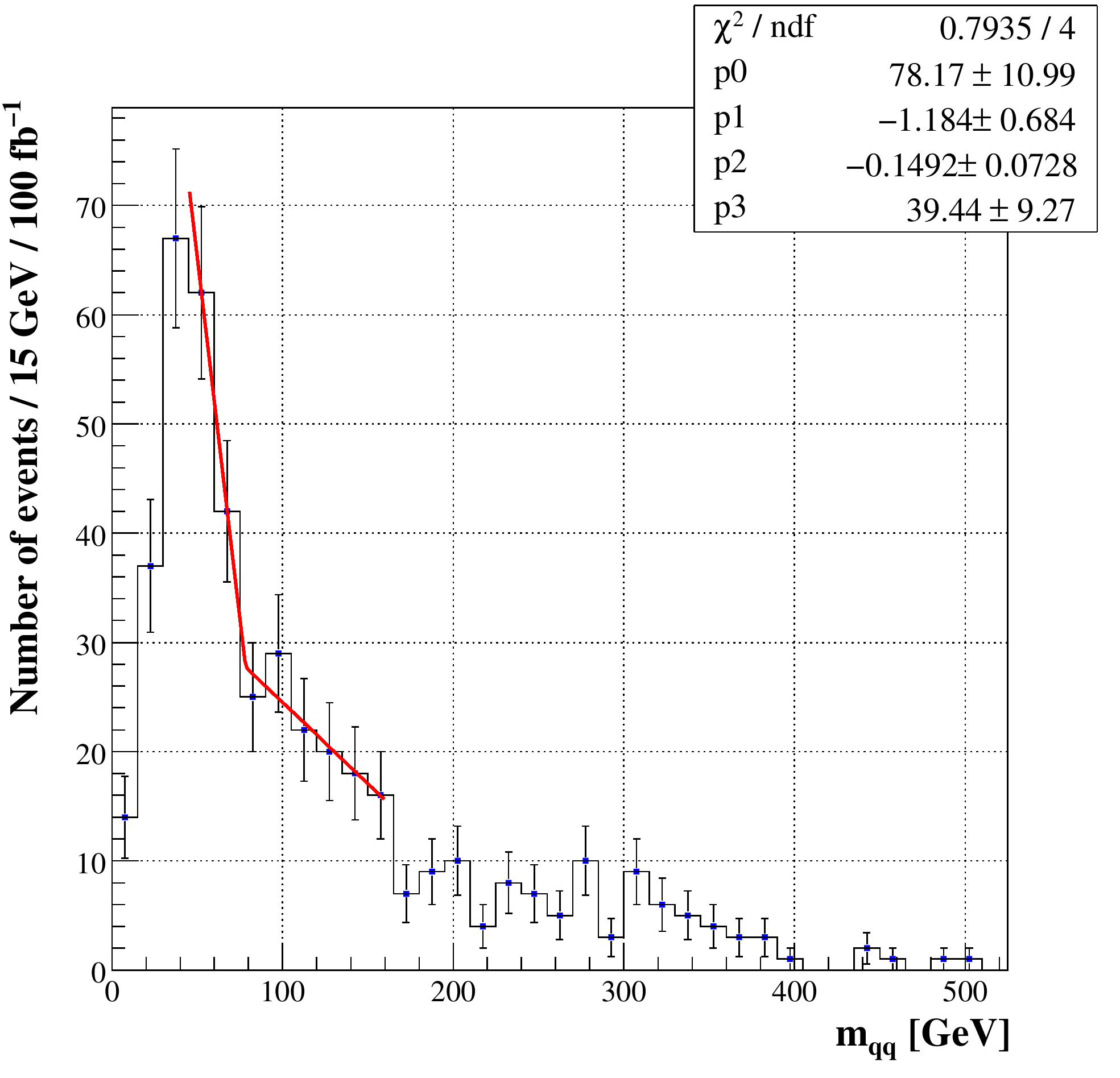} \\
\includegraphics[width=0.40 \textwidth]{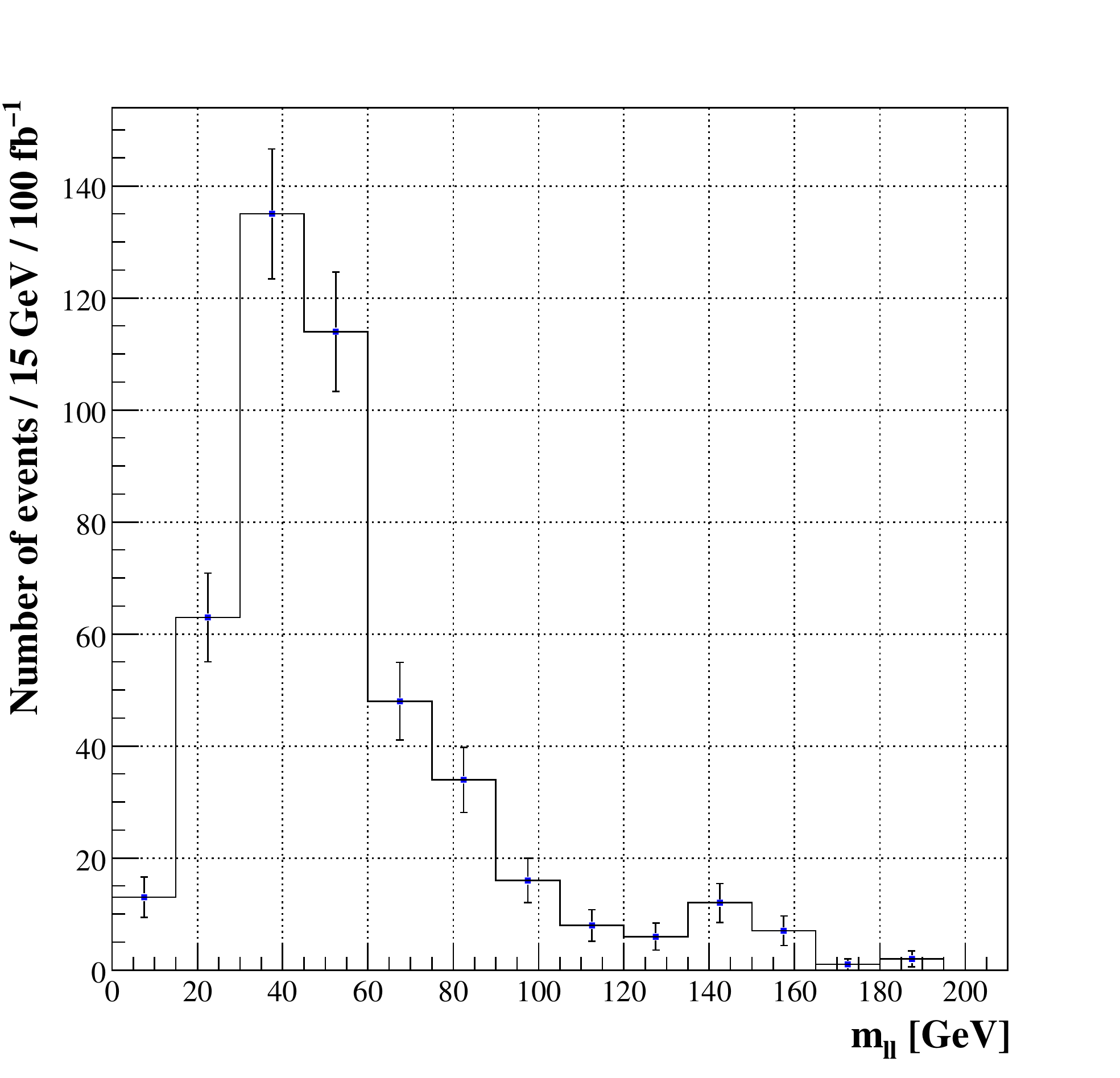}
\caption{{\em Upper-left panel:} invariant mass distribution of the $\ell^+ \ell^-$ pair (step \AGRSchaneulll{},
\agrs{} scenario);
{\em Upper-right panel:} di-jet invariant mass distribution from $\chai$ (step \AGRSchaneuqqll{}, \agrs{} scenario);
{\em Lower panel:} invariant mass distribution of the $\ell^+ \ell^-$ pair (step \AGRSchaneuqqll{},
\agrs{} scenario).}
\label{fig:AGRS-steps23}
\end{center}
\end{figure}

This step is completely analogous to step \GRSchaneulll{} of the \grs{} scenario.
We require two leptons ($e$ or $\mu$) of same flavor and opposite charge, one lepton of different flavor
\footnote{At variance with step \GRSchaneulll{} of the \grs{} scenario, here we also allow this one lepton to be
a hadronically decaying tau in order to have larger statistics.}, and no jets.
To remove possible backgrounds, we also impose the following cuts:
\begin{itemize}
\item[{\bf (a)}] $p_{T 1,2} >$ 20, 10 GeV on the 2 leptons of same flavor;
\item[{\bf (b)}] $p_{T} >$ 10 GeV on the lepton of different flavor;
\item[{\bf (c)}] \met $>$ 50 GeV.
\end{itemize}
The invariant mass distribution constructed with the 2 leptons of same flavor is shown in fig. \ref{fig:AGRS-steps23}
(upper-left panel). Again, the accuracy with which lepton signals are in general reconstructed explains the clear
endpoint structure of the invariant mass distribution at 60 GeV. This endpoint agrees with the expected value
$m_\neuii - m_\neui \doteq 59$ GeV from table \ref{tab:examplefits}.

\subsubsection*{Step \AGRSchaneuqqll{}: \AGRSchaneuqqllTEXT}

We select these events by requiring 2 $q$ jets and 2 leptons ($e$ or $\mu$) of same flavor and opposite charge.
We also impose the following cuts :
\begin{itemize}
\item[{\bf (a)}] $p_{T 1,2} >$ 20, 10 GeV on the 2 jets;
\item[{\bf (b)}] $p_{T 1,2} >$ 20, 10 GeV on the 2 leptons;
\item[{\bf (c)}] \met $>$ 50 GeV.
\end{itemize}
The di-jet invariant mass distribution is reported in fig.
\ref{fig:AGRS-steps23} (upper-right panel). As can be seen by the
length of its tail, the background contribution is quite severe for
this channel even after the above cuts. The fitted endpoint allows
to estimate $m_\chai - m_\neui$ to be 78(11) GeV, whose central
value is 35 $\%$ larger than the true value 58 GeV. So, because of
this systematics, the accuracy in the determination of $m_\chai$
turns out to be not as good as for the other mass determinations in
the \agrs{} scenario.

In addition, we can check $m_\neuii - m_\neui$ as obtained in step \AGRSchaneulll{} by constructing the invariant
mass of the two leptons. Fig. \ref{fig:AGRS-steps23} (lower panel) displays the corresponding distribution,
showing again the quite clear endpoint at 60 GeV, though it is more smeared by backgrounds than the distribution
in step \AGRSchaneulll{}.

\section{Conclusions and Outlook}

We have explored, in the concrete example of Yukawa-unified SUSY GUTs, to which extent a quantitative
determination of the lightest part of the SUSY spectrum is practicable with LHC data. Specifically,
we have considered two representative scenarios, both viable in the light of all existing data,
but characterized by some notable differences in the SUSY spectra (summarized in table
\ref{tab:examplefits}), that arguably can be told apart only via their direct measurement.

We have elaborated a mass-determination strategy based on the observation of kinks in the kinematic
variable $\mtt$, namely the so-called $\mtt$-kink method. This method is especially suitable for our
purposes, as it does not rely on the presence of long decay chains, which are not achievable at the 
LHC within our considered classes of models.

\input Tables/table_massrecap.tex
We have studied in detail our strategy on 100 fb$^{-1}$ of data collected at 14 TeV $p$-on-$p$
collisions, using Pythia 6. We have also simulated detector-level effects via PGS4, with the aim
of conveying a hopefully realistic idea of how large a degradation is expected in real signals.

We were able to determine the masses of the gluino, of the lightest chargino, of the first two
neutralinos for both scenarios considered, and also the mass of the lightest stop for the
scenario where this mass is below the gluino's. Our results are summarized in table
\ref{tab:massrecap}, as obtained through a global fit to the outcomes (masses or functions thereof)
of the various steps described in section \ref{sec:results} for either scenario. The results
in table \ref{tab:massrecap} display a good agreement with the corresponding true values in
table \ref{tab:examplefits}, a statistical error always around 20 GeV and a systematics somewhat
larger for the \agrs{} scenario.

Our worked example of Yukawa-unified SUSY GUTs may offer a useful playground for dealing with
other theories which predict similar patterns of SUSY spectra.

Of course, a number of issues are left open by the present study. First, although we have
been focusing on mass determinations, it is also of crucial importance to determine the spin 
of any SUSY particle produced at the LHC. To this aim, we just note here that the approach of 
ref. \cite{maos}, whereby the invisible particle momenta are reconstructed with $\mtt$ in order 
to determine the mother-particle spin, may be applied to the decay of pair-produced gluinos in 
the scenarios of \cite{AlGuRaS,GRS}, from which the gluino spin might be read off.

Further issues are more specifically concerned with mass measurements and the precision achieved
in their determination.
For example, a first relevant question is that of the luminosity needed for the discovery of either
channel. A rough extrapolation of our results indicates that about 5 to 10 fb$^{-1}$ of data would
be sufficient to obtain a statistical error on the mass determinations of about 50 GeV. This amount 
of data would therefore suffice for a 5$\sigma$ discovery of at least the heavier among the particles 
listed in Table \ref{tab:massrecap} for either scenario.

We note as well that we did not try and exploit here the approach of ref. \cite{cho-kim-kim}, 
that has been proposed as a strategy to make the endpoint structure more prominent, and that might be able 
to reduce the systematic errors in the identification of the endpoint position of the kinematic 
variables considered in our analysis. Pursuing these, more refined strategies could also offer
a handle to address the case of lower needed luminosities, where the signal reach may be increased
by also using more optimized cuts, or allow to implement our strategy to data from Tevatron Run II 
and from the LHC running at center-of-mass energies below the design one.

\acknowledgments

KC and DG warmly acknowledge the CERN Theory Institute and the Galileo
Galilei Institute (Florence, Italy), where initial discussions have
taken place. DG is especially indebted to R.~Franceschini,
M.-H.~Genest, S.~Kraml, F.~Mescia, M.~Pierini, S.~Sekmen and
D~.M.~Straub for valuable feedback. KC, SHI and CBP are supported by
the KRF grants funded by the Korean Government (KRF-2007-341-C00010
and KRF-2008-314-C00064) and KOSEF grant funded by the Korean
Government (No. 2009-0080844). The work of DG is supported by the
DFG Cluster of Excellence `Origin and Structure of the Universe'.

\bibliographystyle{JHEP}
\bibliography{GUT-LHC}

\end{document}

%% file: Tables/table_SUSYprod.tex
\TABLE[t]{
\small
\renewcommand{\arraystretch}{1.2}
\begin{tabular}{|l cc c cc|}
\hline
\multicolumn{6}{|c|}{{\bf SUSY production cross sections}} \\
& \multicolumn{2}{c}{\agrs{} scenario} & & \multicolumn{2}{c|}{\grs{} scenario} \\
& 14 TeV & 10 TeV & & 14 TeV & 10 TeV \\
\hline
total (pb)
& 41 & 17 & & 137 & 57 \\
[0.2cm]
$\tilde\chi_1^\pm \tilde\chi_1^\mp$ (\%)
& 12.7 & 18.7 & & 5.7 & 8.7 \\
$\tilde\chi_2^0 \tilde\chi_1^\pm$ (\%)
& 24.7 & 36.6 & & 11.2 & 17.1 \\
$q \ov q \to \tilde g \tilde g$ (\%)
& 9.1 & 9.7 & & 5.1 & 6.1 \\
$g g \to \tilde g \tilde g$ (\%)
& 53.4 & 34.9 & & 39.7 & 28.9 \\
$g g \to \tilde t_1 \tilde t_1^*$ (\%)
& 0.04 & 0.02 & & 35.2 & 35.1 \\
$f \ov f \to \tilde t_1 \tilde t_1^*$ (\%)
& 0.02 & 0.01 & & 3.0 & 4.1	 \\
\hline
\end{tabular}
\caption{SUSY production cross sections from Pythia 6.42. Decay tables are calculated with SUSYHIT and then fed to Pythia.
Missing entries are meant to be below 0.1 permil.}
\label{tab:cross-sections}
}

%% file: Tables/table_decays.tex
\TABLE[t]{
\small
\renewcommand{\arraystretch}{1.1}
\begin{tabular}{|l c c c|}
\hline
\multicolumn{4}{|c|}{{\bf Main decay modes (\%)}} \\
 & \agrs{} scenario & & \grs{} scenario \\
\hline
$\Ga_{\tilde g}$ (GeV)
& $1.6 \cdot 10^{-5}$ & & $3.3$ \\
[0.1cm]
$\tilde g \to \tilde \chi_2^0 g$
& 0.9 & & -- \\
$\tilde g \to \tilde \chi_2^0 b \ov b$
& 41.7 & & -- \\
$\tilde g \to \tilde \chi_2^0 q \ov q ~~~ (q = u,d,s,c)$
& 0.5 & & -- \\
$\tilde g \to \tilde \chi_1^0 t \ov t$
& 20.1 & & -- \\
$\tilde g \to \tilde \chi_1^0 b \ov b$
& 3.6 & & -- \\
$\tilde g \to \tilde \chi_1^\pm t b$
& 31.5 & & -- \\
$\tilde g \to \tilde t_1^{(*)} t$
& -- & & 100 \\
[0.15cm]
\hline
$\Ga_{\tilde t_1}$ (GeV)
& 25 & & $3.5 \cdot 10^{-3}$ \\
[0.1cm]
$\tilde t_1 \to \tilde g t$
& 93.4 & & -- \\
$\tilde t_1 \to \tilde \chi_1^0 t$
& 5.8 & & -- \\
$\tilde t_1 \to \tilde \chi_1^+ b$
& 0.6 & & 100 \\
[0.15cm]
\hline
$\Ga_{\tilde \chi_2^0}$ (GeV)
& $8.2 \cdot 10^{-11}$ & & $2.6 \cdot 10^{-10}$ \\
[0.1cm]
$\tilde \chi_2^0 \to \tilde \chi_1^0 \gamma$
& 62.2 & & 94.6 \\
$\tilde \chi_2^0 \to \tilde \chi_1^0 b \ov b$
& 31.5 & & 0.8 \\
$\tilde \chi_2^0 \to \tilde \chi_1^0 \ell^+ \ell^- ~~~ (\ell = e, \mu)$
& 0.7 & & 1.0 \\
$\tilde \chi_2^0 \to \tilde \chi_1^0 \tau^+ \tau^-$
& 1.8 & & 0.9 \\
[0.15cm]
\hline
$\Ga_{\tilde \chi_1^\pm}$ (GeV)
& $1.8 \cdot 10^{-11}$ & & $1.2 \cdot 10^{-11}$ \\
[0.1cm]
$\tilde \chi_1^\pm \to \tilde \chi_1^0 \tau \nu_\tau$
& 91.0 & & 39.5 \\
$\tilde \chi_1^\pm \to \tilde \chi_1^0 \ell \nu_\ell ~~~ (\ell = e, \mu)$
& 4.7 & & 42.7 \\
$\tilde \chi_1^\pm \to \tilde \chi_1^0 ud / cs$
& 4.3 & & 17.8 \\
\hline
\end{tabular}
\caption{Main decay modes in \% for $\tilde g$, $\tilde t_1$, $\tilde \chi_2^0$ and $\chai$, calculated with SUSYHIT.}
\label{tab:decay-modes}
}

%% file: Tables/table_strategyrecap.tex
\TABLE[t]{
\small
\renewcommand{\arraystretch}{1.3}
\begin{tabular}{|c c c c c|}
\hline
 & \bf Step & \bf Event trigger & \bf Event variable & \bf Allows to determine \\
\hline
\hline
\multirow{6}{*}{\begin{sideways} \bf \phantom{,} \grs{} scenario\end{sideways}} 
  & \GRSgogo & $\tilde g \tilde g \to 4 b + 2 W ~ (+ 2 \ell) + \sla{p}_T$ & $\mtt$ & $m_{\tilde g}$ and $m_\chai$\\
[-0.2cm]
& & & {\footnotesize (with $2 \ell \in$ $\sla{p}_T$)} & \\
& & & & \\
[-0.4cm]
& \GRStt & $\stopi \stopi \to 2 b + 4 q + \sla{p}_T$ & $m_{T,bqq}$ & $m_\stopi - m_\neui$ \\
& '' & '' & $m_{T,qq}$ & $m_\chai - m_\neui$ \\
& & & & \\
[-0.4cm]
& \GRSchaneulll & $\chai \neuii \to \ell^+ \ell^- \ell' + \sla{p}_T$ & $m_{\ell \ell}$ & $m_\neuii - m_\neui$ \\
[0.1cm]
\hline
\hline
\multirow{6}{*}{\begin{sideways} \bf \phantom{,} \agrs{} scenario\end{sideways}} 
  & \AGRSgogo & $\tilde g \tilde g \to 4 b ~ (+ 2 \gamma) + \sla{p}_T$ & $\mtt$ & $m_{\tilde g}$ and $m_\neuii$\\
[-0.2cm]
& & & {\footnotesize (with $2 \gamma \in$ $\sla{p}_T$)} & \\
& & & & \\
[-0.4cm]
& \AGRSchaneulll & $\chai \neuii \to \ell^+ \ell^- + \ell' + \sla p_T$ & $m_{\ell \ell}$ & $m_\neuii - m_\neui$ \\
& & & & \\
[-0.4cm]
& \AGRSchaneuqqll & $\chai \neuii \to 2 q + 2 \ell + \sla p_T$ & $m_{q q}$ & $m_\chai - m_\neui$ \\
& '' & '' & $m_{\ell \ell}$ & $m_\neuii - m_\neui$ \\
[0.1cm]
\hline
\end{tabular}
\caption{Overview of our mass-determination strategy.}
\label{tab:strategyrecap}
}

%% file: Tables/table_massrecap.tex
\TABLE[ht]{
\small
\renewcommand{\arraystretch}{1.3}
\begin{tabular}{|c c c|}
\hline
& {\bf Quantity} & {\bf Result (GeV)} \\
\hline
\hline
\multirow{5}{*}{\begin{sideways} \bf \grs{} scenario\end{sideways}} 
& $m_\go$ & $395 \pm 16$ \\
& $m_\chai$ & $109 \pm 17$ \\
& $m_\neui$ & $57 \pm 17$ \\
& $m_\neuii$ & $107 \pm 18$ \\
& $m_\stopi$ & $206 \pm 17$ \\
\hline
\hline
\multirow{4}{*}{\begin{sideways} \bf \agrs{} scenario\phantom{x}\end{sideways}} 
& $m_\go$ & $456 \pm 15$ \\
& $m_\chai$ & $144 \pm 20$ \\
& $m_\neui$ & $66 \pm 16$ \\
& $m_\neuii$ & $126 \pm 16$ \\
[0.1cm]
\hline
\end{tabular}
\label{tab:massrecap}
\caption{Mass determinations within our strategy, to be compared with table \ref{tab:examplefits}.}
}